\documentclass[12pt,draftclsnofoot,onecolumn]{IEEEtran}
\ifCLASSINFOpdf
\else
\fi

\usepackage{graphicx}
\usepackage[cmex10]{amsmath}
\usepackage{hyperref}
\usepackage{amsfonts}
\usepackage{dblfloatfix}
\usepackage{cite}
\usepackage{dsfont}
\usepackage{bm}
\usepackage{amssymb}
\usepackage{array}
\usepackage{algorithm}
\usepackage{algorithmic}
\usepackage{subfigure}
\usepackage{graphics}
\usepackage{epsfig}
\usepackage{epstopdf}
\usepackage{float}
\usepackage{extarrows}
\usepackage{makecell}
\usepackage{booktabs}
\usepackage{xcolor}

\newtheorem{remark}{Remark}

\begin{document}
%\abovedisplayskip=0.5pt
%\belowdisplayskip=0.5pt
%\allowdisplaybreaks

\title{Message Passing-Based Joint User Activity Detection and Channel Estimation for Temporally-Correlated Massive Access}
%
%\author{
%\IEEEauthorblockN{Weifeng~Zhu, Meixia~Tao, and Yunfeng~Guan} \\
%\IEEEauthorblockA{Department of Electronic Engineering, Shanghai Jiao Tong University, Shanghai, China \\
%Emails: \{wf.zhu, mxtao, yfguan69\}@sjtu.edu.cn}
%}
\author{
Weifeng~Zhu,~\IEEEmembership{Member,~IEEE}, Meixia~Tao,~\IEEEmembership{Fellow,~IEEE}, Xiaojun~Yuan,~\IEEEmembership{Senior Member,~IEEE}, and Yunfeng~Guan

\thanks{This paper was presented in part at the IEEE International Conference of Communications (ICC) 2021 \cite{Zhu_2021_ICC}.}
\thanks{W. Zhu, M. Tao, and Y. Guan are with the Department of Electronic Engineering, Shanghai Jiao Tong University, Shanghai 200240,
China (e-mail: \{wf.zhu, mxtao, yfguan69\}@sjtu.edu.cn).}
\thanks{X. Yuan is with the Center for Intelligent Networking and Communication (CINC), University of Electronic Science and Technology of China, Chengdu 610000, China (e-mail: xjyuan@uestc.edu.cn).}
%\thanks{ is with the Department of Electronic Engineering, Shanghai Jiao Tong University, Shanghai 200240,
%China  (e-mail: @sjtu.edu.cn).}
}
\maketitle

\vspace{-1.4cm}

\begin{abstract}

This paper studies the user activity detection and channel estimation problem in a temporally-correlated massive access system where a very large number of users communicate with a base station sporadically and each user once activated can transmit with a large probability over multiple consecutive frames.
We formulate the problem as a dynamic compressed sensing (DCS) problem to exploit both the sparsity and the temporal correlation of user activity.
By leveraging the hybrid generalized approximate message passing (HyGAMP) framework, we design a computationally efficient algorithm, HyGAMP-DCS, to solve this problem.
In contrast to only exploit the historical estimations, the proposed algorithm performs bidirectional message passing between the neighboring frames for activity likelihood update to fully exploit the temporally-correlated user activities.
Furthermore, we develop an expectation maximization HyGAMP-DCS (EM-HyGAMP-DCS) algorithm to adaptively learn the hyperparameters during the estimation procedure when the system statistics are unknown.
In particular, we propose to utilize the analysis tool of state evolution to find the appropriate hyperparameter initialization of EM-HyGAMP-DCS.
Simulation results demonstrate that our proposed algorithms can significantly improve the user activity detection accuracy and reduce the channel estimation error.

\end{abstract}
\vspace{-0.5cm}
\begin{IEEEkeywords}
Temporally-correlated massive access, user activity detection, channel estimation, hybrid generalized approximate message passing (HyGAMP), expectation maximization (EM).
\end{IEEEkeywords}

\section{Introduction}\label{sec:introduction}

%To support massive devices connectivity for the Internet of Things (IoT) application, the massive machine-type communications (mMTC) has been defined as one of the main use cases of the fifth-generation (5G) cellular networks \cite{Chen_2020_JSAC}.
%The key features of the mMTC are usually characterized by a massive number of devices, sporadic transmissions and short packet size \cite{Bock_2016_CM}.
%The main challenge of mMTC is to design scalable and efficient multiple-access schemes with low latency. While the conventional grant-based random access schemes usually suffer excessive delay and signaling overhead, the grant-free random access scheme is considered as the promising solution with minimum control overhead for the massive access systems \cite{Liu_2018_SPM}.

Massive machine-type communications (mMTC) is one of the main use cases of the fifth-generation (5G) cellular networks, for Internet of Things (IoT) applications \cite{Chen_2021_JSAC}.
It aims to provide wireless connectivity to a massive number of IoT devices, whose traffic is typically sporadic \cite{Bock_2016_CM, Liu_2018_SPM}.
The main technical challenge of mMTC is to design scalable, efficient, and low-latency multiple-access schemes. Due to the massive number of device, the conventional grant-based random access schemes suffer from excessive delay and signaling overhead, then the grant-free random access scheme is considered as the promising solution to decrease the access delay and reduce the control overhead for coordination \cite{Liu_2018_SPM}.

In the grant-free random access protocol, a unique pilot sequence for identification and channel estimation is pre-allocated to each device. Note that, due to the massive number of the IoT devices but the limited coherence time interval, the pilot sequences are usually non-orthogonal. When one device is activated, it directly transmits its dedicated pilot followed by data without waiting for the grant from the base station (BS). Meanwhile, the BS can perform joint user activity detection and channel estimation based on the received signals. Given the sporadic traffic generated from IoT devices, the joint user activity detection and channel estimation problem by nature can be cast into a large-scale sparse signal recovery problem, which can be reliably solved by the compressed sensing (CS) techniques \cite{Liu_2018_SPM}.

%In massive access, the users usually have a high probability to continue transmitting information to the BS in multiple consecutive frames when being activated by bursty events. As such, the inherent temporal correlation can be exploited for higher-quality user activity detection and channel estimation.
In many practical IoT systems, once a device is activated, it often transmits continuously over multiple frames. This suggests that the device activities are often temporally-correlated.
%By exploiting both the temporal correlation and sparsity in user activities, this paper aims to propose a new algorithm for joint user activity and channel estimation which significantly outperforms the state-of-the art approaches.
%Though we can sequentially perform user activity detection and channel estimation in each frame with the use of the historical estimation results, this approach ignores statistical relationship between the user activity in the current frame and those in the following frames and usually leads to several detection error accumulation.
%\textcolor[rgb]{0.00,0.07,1.00}{
To exploit the temporally-correlated user activity, we are motivated to detect the active users and estimate their channels in multiple consecutive frames jointly. The joint user activity detection and channel estimation problem can be formulated from the \emph{dynamic compressed sensing} (DCS) perspective \cite{Angle_2009_ICDSP, Ziniel_2013_TSP}, which takes both the sporadic user activity and the temporal correlation of user activities into account. Based on the probabilistic and graphical model of the signals, the joint user activity detection and channel estimation can be realized by the standard message passing (MP) algorithm \cite{Ksch_2001_TIT}. To simplify the implementation of MP in large-scale systems, we propose to leverage the hybrid generalized approximate message passing (HyGAMP) \cite{Rangan_2017_TSP} framework which employs a hybrid of AMP and MP for the graphical model. Considering that the perfect system statistics are usually unknown, the expectation maximization (EM) technique is incorporated with the algorithm to adaptively learn the hyperparameters in the estimation procedure. Compared with the existing DCS-based algorithms, we show that the proposed algorithms can significantly improve the user activity detection accuracy and reduce the channel estimation error.
%}

\subsection{Related Works}

To accomplish joint user activity detection and channel estimation for massive access, a variety of advanced sparse signal recovery algorithms have been proposed for different wireless communication systems.
For the single-cell scenario, the works \cite{Schepker_2013_ISWCS, Wunder_2015_GCW} utilize the standard CS algorithms of orthogonal matching pursuit (OMP) and basis pursuit denoising (BPDN). However, these two kinds of algorithms usually suffer high computational complexity due to the extremely large amount of devices.
%The computationally efficient AMP algorithm becomes an attractive approach to realize JUADCE with improved performance, which can also exploit the system statistics.
Then the works \cite{Chen_2018_TSP, Liu_2018_TSP} propose the computationally efficient AMP algorithms with Bayesian denoiser for user activity detection. In the work \cite{RB_2021_WCL}, the message-scheduling generalized AMP algorithm is developed to further reduce the computational complexity without degrading the detection and estimation performance. The AMP-based algorithms are also extended to perform data detection along with the activity detection in \cite{Senel_2018_TCOM, Jiang_2020_TWC}. Besides the AMP-based algorithms, covariance matching pursuit \cite{Fengler_2021_TIT, Chen_2021_TIT}, dimension reduction-based optimization \cite{Shao_2020_TSP} and deep learning methods \cite{Zhang_2019_TVT, Shao_2021_JSAC, Zhu_2021_TWC} are also investigated for performance enhancement. In particular, the covariance matching pursuit algorithm \cite{Fengler_2021_TIT, Chen_2021_TIT} and dimension reduced-based optimization algorithm \cite{Shao_2020_TSP} are especially designed for the massive multiple-input multiple-out (MIMO) systems, which can significantly outperform the conventional CS-based methods. The deep learning methods \cite{Zhang_2019_TVT, Shao_2021_JSAC, Zhu_2021_TWC} employ the artificial neural networks and learn the network parameters based on the pre-collected data, which are able to outperform the traditional hand-designed algorithms and to enjoy low computational complexity. Moreover, a sparsity-constrained method is proposed for the non-ideal scenario where different users have unknown frequency offsets in \cite{Li_2019_TWC}. In the multi-cell systems, the cooperative user activity detection and channel estimation is studied in \cite{Chen_2019_TWC, Ke_2021_JSAC} based on the AMP-based algorithms. Recently, unsourced random access as a new random access protocol is also studied in \cite{Poly_2017_ISIT, Fengler_2021_TIT}, where all users share a common codebook and the BS only needs to detect the transmit codewords.

%Moreover, the work \cite{Li_2019_TWC} proposes the sparsity-constrained method for the non-ideal scenario where different users have unknown frequency offsets.
%\textcolor[rgb]{0.00,0.07,1.00}{
The aforementioned works all detect the user activity in each frame individually by assuming the user activity is independent for each frame.
Due to the low-rate transmission, the information transmission of one active user may occupy multiple consecutive frames. Such temporal correlation of user activity can be exploited for performance enhancement.
%}
%the temporal correlation in the user activity can also be considered for user activity detection \cite{Wang_2016_CL, Du_2017_JSAC, Wang_2021_arxiv, Jiang_2021_TWC}. %which is reformulated as a DCS problem.
Assuming the user sparsity level and the channel state information (CSI) are available, the work \cite{Wang_2016_CL} proposes a DCS-based algorithm where the set of the detected active user in the last slot is used as the initial set to be detected in the current slot. By adaptively exploiting the prior support based on the corresponding support quality information, the work \cite{Du_2017_JSAC} proposes a prior-information-aided adaptive subspace pursuit (PIA-ASP) algorithm, which can always outperform the DCS-based algorithm in \cite{Wang_2016_CL}.
%To fully extract the temporal correlation over multiple snapshots, a side-information aided block OMP (SIA-BOMP) algorithm is also proposed in \cite{Cui_2020_TWC}.
These algorithms include the matrix inversion calculation in each iteration, and thus suffer high computational complexity in the large-scale problems. Moreover, both the works \cite{Wang_2016_CL, Du_2017_JSAC} assume that the CSI is perfectly known in advance, which is unrealistic in massive access. By leveraging the system statistics, the authors in \cite{Jiang_2021_TWC} propose a sequential AMP (S-AMP) algorithm to sequentially perform user activity detection and channel estimation in each frame.
Then the work \cite{Wang_2021_ISIT} proposes to extract the side information (SI) from the estimation results in the previous frame to enhance the user activity detection performance in the current frame based on the AMP framework.
Different from \cite{Wang_2021_ISIT} which only utilizes the historical estimation results as SI, our previous work \cite{Zhu_2022_WCL} proposes to extract the double-sided information by considering the estimation from the next frame as well for further exploiting the temporal correlation.
%Note that these prior works \cite{Wang_2016_CL, Du_2017_JSAC, Wang_2021_ISIT, Jiang_2021_TWC} all only exploit the temporal correlation by using the historical estimation information.
%To further exploit the temporal correlation, the work \cite{Zhu_2022_WCL} proposes to additionally utilize the estimation information in the next frame based on the AMP algorithm with SI, which, however, doubles the computational cost.

%\textcolor[rgb]{0.00,0.07,1.00}{
%This paper proposes to perform joint activity detection and channel estimation in multiple consecutive frames simultaneously, where the statistical correlation between the activities in these consecutive frames can be fully exploited to provide significant performance improvement.
%Moreover, we propose to adaptively learn the parameters of the system statistics based on the EM technique, while the works \cite{Wang_2021_arxiv, Jiang_2021_TWC} assume perfect system statistics being available. To ensure the HyGAMP-based algorithm incorporated with EM technique converges to the satisfied point and have a fast convergence, we then propose to employ performance analysis tool of the state evolution (SE) to help find the appropriate hyperparameter initialization.
%The work \cite{Zhu_2022_WCL} proposes to additionally utilize the estimation information in the next frame to further exploit the temporal correlation and it also performs frame-by-frame activity detection as the works \cite{Wang_2016_CL, Du_2017_JSAC, Wang_2021_ISIT, Jiang_2021_TWC} do.
Compared with these prior works \cite{Wang_2016_CL, Du_2017_JSAC, Wang_2021_ISIT, Jiang_2021_TWC, Zhu_2022_WCL} that exploit the temporal correlation for use activity detection and channel estimation, this paper differs mainly in the following two aspects. First, while all the exiting works perform activity detection in a frame-by-frame manner, (i.e., the user activity is still detected sequentially frame by frame, though the temporal correlation among adjacent frames has been exploited), this paper performs multi-frame activity detection in a block-by-block manner. Second, this paper proposes to adaptively learn the parameters of the system statistics by using the EM technique, while the previous algorithms either do not consider the system statistics \cite{Wang_2016_CL, Du_2017_JSAC} or assume the system statistics are perfectly known \cite{Wang_2021_ISIT, Jiang_2021_TWC, Zhu_2022_WCL}.

\subsection{Main Contributions}
In this work, we consider the problem of joint user activity detection and channel estimation in multiple consecutive frames for the temporally-correlated massive access systems.
%\textcolor[rgb]{0.00,0.07,1.00}{In contrast to perform frame-by-frame detection, we  multi-frame activity detection problem, where .}
%
%Independent to the work \cite{Jiang_2021_TWC}, we also propose to describe the activity evolution of each user with the first-order Markov chain. Then we also account the system statistics to design a scalable detection and estimation algorithm called HyGAMP-DCS with outstanding recoverability and efficient computational complexity. Since the perfect system statistics are usually hard to be acquired, the EM algorithm is incorporated in the HyGAMP-DCS algorithm to adaptively update the hyperparameters.
%
The main contributions and distinctions of this work are summarized as follows:
\begin{enumerate}
  \item We propose to perform the user activity detection and channel estimation jointly in multiple consecutive frames and formulate it as a DCS problem. Based on the probabilistic model, not only the sparse activity pattern is considered in the problem, but also the statistical relationships of the activities in these consecutive frames are exploited.
  %\item This paper proposes to reformulate the problem in the probabilistic perspective, which uses a steady first-order Markov chain to characterize the user activity evolution  and takes advantage of the channel statistics. Then the powerful Bayesian inference methods can be employed to solve the problem.
  \item %\textcolor[rgb]{0.00,0.07,1.00}{
        In contrast to only making use of the historical estimations in \cite{Jiang_2021_TWC, Wang_2021_ISIT}, this paper proposes a HyGAMP-DCS algorithm to fully exploit the temporally-correlated user activities in the multiple consecutive frames. The HyGAMP-DCS algorithm combines the computationally efficient GAMP algorithm for channel estimation and the standard MP algorithm for the activity likelihood update. In particular, the activity likelihood values in each frame are updated by aggregating both the forward messages from the previous frame and backward messages from the next frame at each iteration.
      %Specifically, HyGAMP-DCS can be divided in to two parts of GAMP and MP. The GAMP part performs channel estimation computational efficient GAMP  where estimated activity likelihoods between the adjacent frames are exchanged for refinement.  is designed to perform activity detection and channel estimation in multiple consecutive frames jointly. Specifically, both the activities and the channels in each frame are simultaneously refined in each iteration by exchanging the useful information between the adjacent frames.
        Numerical results show that HyGAMP-DCS can achieve superior performance to the existing DCS-based algorithms \cite{Wang_2016_CL,Du_2017_JSAC,Jiang_2021_TWC,Wang_2021_ISIT} thanks to the bidirectional message propagation.
  \item %\textcolor[rgb]{0.00,0.07,1.00}{
      To make the proposed algorithm applicable for the practical system with imperfect system statistics, this paper incorporates the EM algorithm in HyGAMP-DCS to adaptively learn the hyperparameters during the estimation procedure.
      Compared with the traditional EM-AMP algorithm \cite{Vila_2013_TSP} for the CS problem, the proposed EM-HyGAMP-DCS algorithm additionally learns the statistical dependencies between the activities in the neighboring frames.
      %Compared with the existing algorithms in \cite{Jiang_2021_TWC, Wang_2021_ISIT, Zhu_2022_WCL}, the proposed EM-based HyGAMP-DCS (EM-HyGAMP-DCS) algorithm can still achieve favorable performance without perfect system statistics.
      Simulation results demonstrate that EM-HyGAMP-DCS realizes very similar performance to HyGAMP-DCS which requires the perfect system statistics.
  \item %\textcolor[rgb]{0.00,0.07,1.00}{
        We provide the performance and complexity analysis of the proposed algorithms.
        We first introduce the state evolution (SE) to predict the asymptotic performance of HyGAMP-DCS and EM-HyGAMP-DCS. In particular, we also point out that the SE is quite essential to find the appropriate hyperparameter initialization for EM-HyGAMP-DCS, which is validated by the numerical results.
        Then the computational complexity comparison with the state-of-the-art methods is given to illustrate the computational efficiency of our proposed algorithms.
\end{enumerate}

\subsection{Organizations and Notations}

The remaining part of this paper is organized as follows.
Section \ref{sec:system_model} introduces the system model of temporally-correlated massive access.
In Section \ref{sec:problem_formulation}, we present the probabilistic model of the considered system and then introduce the Bayesian inference methods.
In Section \ref{sec:HyGAMP-DCS}, the HyGAMP-DCS algorithm is proposed, then the EM algorithm is employed for hyperparameter update in Section \ref{sec:EM}.
Next, we analyze the performance and the computational complexity of the proposed algorithms in Section \ref{sec:performance_analysis}.
The simulation results of the proposed algorithms are shown in Section \ref{sec:simulation}.
Finally, we conclude this paper in Section \ref{sec:conclusion}.

In this paper, upper-case and lower-case letters denote random variables and their realizations, respectively.
Letters $\mathbf{x}$, $\mathbf{X}$ denote vector and matrix, respectively.
Superscripts $(\cdot)^T, (\cdot)^*$ denote transpose and conjugate, respectively.
Further, $\mathbb{E}[\cdot]$ and $\mathbb{V}[\cdot]$ denotes the expectation operation and variance operation, respectively; $|\cdot|$ denotes the magnitude of a variable.
In addition, a random vector $\mathbf{x} \in \mathbb{C}^{M \times 1}$ drawn from the complex Gaussian distribution with mean $\mathbf{x}_0 \in \mathbb{C}^{M \times 1}$ and covariance matrix $\pmb{\Sigma} \in \mathbb{C}^{M \times M}$ is characterized by the probability density function \textcolor[rgb]{0.00,0.07,1.00}{$\mathcal{CN}(\mathbf{x};\mathbf{x}_0,\pmb{\Sigma}) = \frac{\exp\left(-(\mathbf{x}-\mathbf{x}_0)^H\pmb{\Sigma}^{-1}(\mathbf{x}-\mathbf{x}_0)\right)}{\pi^{M}|\pmb{\Sigma}|}$}.

\section{System Model}\label{sec:system_model}

\begin{figure}[t]
  \centering
  \includegraphics[width=.45\textwidth]{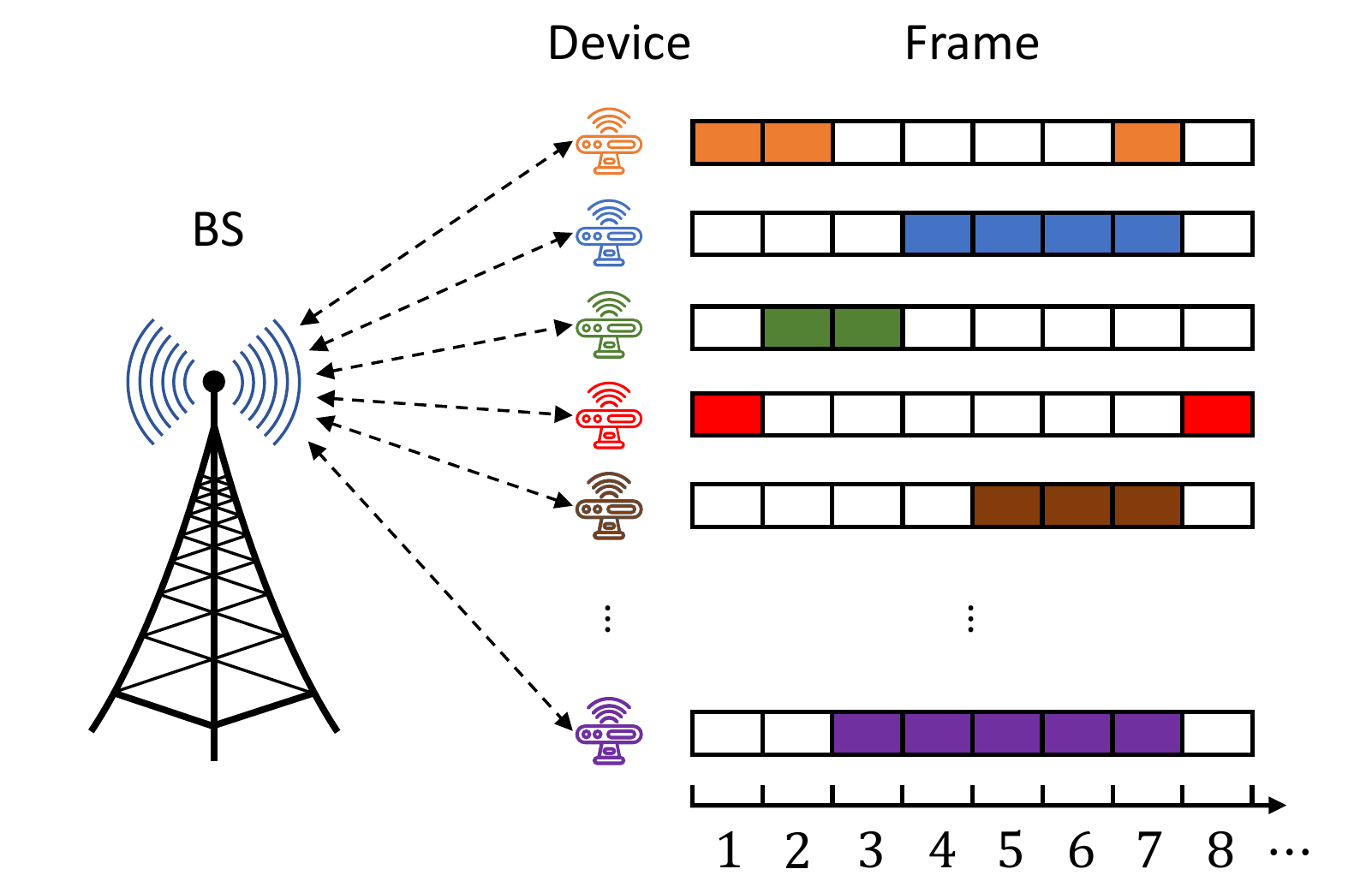}
  \vspace{-0.8cm}
  \caption{Illustration of a temporally-correlated massive access system}\label{Fig:sm}
  \vspace{-0.5cm}
\end{figure}

We consider a single-cell massive access system where a large number, denoted as $N$, of users communicate to a common BS under the grant-free mechanism. The BS and users are all equipped with single antenna. We assume that the transmit signals of all users are perfectly synchronized at the BS. The block fading channel model is adopted, i.e., the channel coefficients on each communication link keep unchanged within a transmission frame but vary in different frames. Due to the sporadic traffic of IoT applications, only a small portion of users are active for uplink transmission while others keep silent at each time frame.

\subsection{Temporal Correlation of the User Activity}\label{subsec:UAM}
As shown in Fig. \ref{Fig:sm}, once a user is activated, its data transmission can occupy multiple consecutive frames. As such, there exist temporal correlations of the user activities in neighboring frames.

Similar to \cite{Jiang_2021_TWC,Wang_2021_ISIT,Zhu_2022_WCL,Ziniel_2013_TSP}, the dynamic user activity over frames is modeled as a stationary first-order Markov chain with two discrete states. The Markov chains on different devices are assumed to independent and identically distributed (i.i.d.). Let $\lambda_{n,t} \in \{0,1\}$ indicate whether user $n \in \{1,2,\dots,N\}$ is active or not in the $t$th frame. The steady state probability is given by $\text{Pr}(\lambda_{n,t} = 1) = p_a$, where $p_a$ denotes the active probability of each user $n$ in frame $t$.  The transition probability matrix is defined as
\begin{equation}\label{equ:PTM}
    \mathbf{P} = \left[ \begin{array}{cc}
                          p_{00} & p_{10} \\
                          p_{01} & p_{11}
                        \end{array} \right],
\end{equation}
where $p_{10} = \text{Pr}\{\lambda_{n,t}=0|\lambda_{n,t-1}=1\}$ and other three transition probabilities are defined similarly.
%process\footnote{The steady process means that $\mathbf{P}\left[
%                                                                                                                      \begin{array}{c}
%                                                                                                                        1-p_a \\
%                                                                                                                        p_a
%                                                                                                                      \end{array}\right] = \left[\begin{array}{c}
%                                                                                                                                             1-p_a \\
%                                                                                                                                             p_a
%                                                                                                                                           \end{array}\right]$ and the expression of $p_{01}$ can be acquired as $p_{01} = p_a p_{10}/(1-p_a)$ by solving the eigenvalue problem.}
By the steady-state assumption, the Markov chain can be completely characterized by two parameters $p_a$ and $p_{10}$.
The other three transition probabilities can be easily obtained as $p_{01} = p_a p_{10}/(1-p_a)$, $p_{00} = 1 - p_{01}$, and $p_{11} = 1 - p_{10}$.
Note that $1/p_{10}$ can be considered as the expected number of the consecutive frames occupied by the data transmission of one user, which means that smaller $p_{10}$ leads to stronger temporal correlation of the user activity over frames.
Although the temporal correlation can be described more precisely by higher-order Markov chains, we consider the first-order Markov chain here for simplicity.

\subsection{Multi-Frame Dynamic Signal Model}

In order to exploit the temporally correlated user activities, we consider $T$ consecutive frames for signal transmission with $T\ge2$.
In each frame $t \in \{1,2,\dots,T\}$, the transmission is divided into two phases, the pilot phase and data phase. Each user $n$ is pre-assigned a unique pilot sequence $\mathbf{a}_n=[a_{n,1},a_{n,2},\dots,a_{n,L}]^T \in \mathbb{C}^{L\times1}$ with length $L$. The BS performs user identification and channel estimation based on the detected pilot sequences in the pilot phase and then decodes the data in the data phase. In this work, we concentrate on the joint user activity detection and channel estimation problem in the pilot phase. Due to the limit of orthogonal resources, the pilot length $L$ is assumed to be much smaller than the number of users $N$, i.e., $L \ll N$, so that the pilot sequences of different users cannot be mutually orthogonal. The pilot sequences in this work are generated from the i.i.d. complex Gaussian distributions with zero mean and variance of $1/L$, and then scaled to have unit power.

The channel in each frame is assumed to satisfy independent Rayleigh distribution. We denote the channel coefficient between user $n$ and the BS in the $t$th frame as $h_{n,t} = \sqrt{\beta_{n}} g_{n,t}$, where $\beta_{n} = P_n\gamma_n$ is the effective large-scale fading component decided by the transmit power $P_n$ and the large-scale attenuation $\gamma_n$, $g_{n,t} \in \mathbb{C}$ is the small-scale fading coefficient following i.i.d. circularly symmetric complex Gaussian distributions with zero mean and unit variance, i.e., $g_{n,t} \sim \mathcal{CN}(0,1)$. We adopt the power control strategy in \cite{Senel_2018_TCOM}, so that the effective large-scale fading component of each user is the same, i.e., $\beta_{n} = \beta, \forall n$. Then the received pilot signals at the BS in continuous $T$ frames can be written as
%\begin{align}\label{equ:rece_sig_t}
%    \mathbf{y}_{t} &= \sum_{n=1}^{N}\lambda_{n,t} h_{n,t} \mathbf{a}_n + \mathbf{w}_{t}, \notag \\\
%    \quad              &= \mathbf{A}\left(\mathbf{\lambda}_t\odot\mathbf{h}_{t}\right) + \mathbf{w}_{t}, \notag \\
%    \quad              &\triangleq \mathbf{A}\mathbf{x}_{t} + \mathbf{w}_{t},
%\end{align}
%where $\mathbf{A} = [\mathbf{a}_1,\mathbf{a}_2,\dots,\mathbf{a}_N] \in \mathbb{C}^{L \times N}$ is the pilot matrix of all users; $\mathbf{\lambda}_t=[\lambda_{1,t},\lambda_{2,t},\dots,\lambda_{N,t}]^T \in \mathbb{C}^{N \times 1}$ is the activity vector;
%$\mathbf{h}_t = [h_{1,t},h_{2,t},\dots,h_{N,t}]^T \in \mathbb{C}^{N \times 1}$ is the channel vector;
%$\mathbf{x}_t = [x_{1,t},x_{2,t},\dots,x_{N,t}]^T \in \mathbb{C}^{N \times 1}$ is the effective channel vector which is the Hadamand product of $\mathbf{\lambda}_t$ and $\mathbf{h}_t$;
%$\mathbf{w}_t \in \mathbb{C}^{L \times 1}$ is the effective additive noise vector following i.i.d. complex Gaussian distribution whose variance $\sigma_w$ is normalized by the transmit power of users. Then received signal in the $T$ continuous frames can be obtained as
\begin{align}\label{equ:rece_sig_T}
\abovedisplayskip=1pt
\belowdisplayskip=1pt
    \mathbf{Y} &= \mathbf{A} \left(\mathbf{\Lambda} \odot \mathbf{H}\right) + \mathbf{W} \triangleq \mathbf{A}\mathbf{X} + \mathbf{W},
\end{align}
where $\mathbf{A} = [\mathbf{a}_1,\mathbf{a}_2,\dots,\mathbf{a}_N] \in \mathbb{C}^{L \times N}$ is the pilot matrix of all users;
$\mathbf{\Lambda} = [\pmb{\lambda}_1, \pmb{\lambda}_2, \dots, \pmb{\lambda}_T] \in \mathbb{C}^{N \times T}$ is the activity matrix with $\pmb{\lambda}_t=[\lambda_{1,t},\lambda_{2,t},\dots,\lambda_{N,t}]^T \in \mathbb{C}^{N \times 1}$;
$\mathbf{H} = [\mathbf{h}_1,\mathbf{h}_2,\dots,\mathbf{h}_T] \in \mathbb{C}^{N \times T}$ is the channel matrix with $\mathbf{h}_t = [h_{1,t},h_{2,t},\dots,h_{N,t}]^T \in \mathbb{C}^{N \times 1}$;
$\mathbf{X} = [\mathbf{x}_1,\mathbf{x}_2,\dots,\mathbf{x}_T] \in \mathbb{C}^{N \times T}$ is the effective channel matrix defined as the Hadamard product of $\mathbf{\Lambda}$ and $\mathbf{H}$ with each entry $x_{n,t}=\lambda_{n,t} h_{n,t}$;
$\mathbf{W} = [\mathbf{w}_1,\mathbf{w}_2,\dots,\mathbf{w}_T] \in \mathbb{C}^{L \times T}$ is the additive white complex Gaussian noise (AWGN) matrix where each element has zero mean and variance $\sigma^2_w$.

Our objective is to jointly detect the active users and estimate their channels by recovering $\mathbf{X}$ from the received signal $\mathbf{Y}$ given the pilot matrix $\mathbf{A}$.
%The problem can be regarded as an underdetermined linear inverse problem in the multiple measurement vector (MMV) form.
%Due to the sparse user activity pattern, each column of the effective channel $\mathbf{X}$, i.e., $\mathbf{x}_{t}$, only has a small portion of non-zero entries, and thus the powerful CS-based algorithms can be readily applied. However, it is noticed that the positions of the non-zeros entries in each $\mathbf{x}_t$ slowly changes with $t$ due to the temporally-correlated user activity.
This underdetermined linear inverse problem can be referred to as the \emph{dynamic compressed sensing} (DCS) problem, where the support of the vector $\mathbf{x}_t$ slowly changes with $t$.
%In this work, we propose to exploit the temporal correlation of user activity in the DCS problem to enhance the performance, which will be detailed in the following sections.
%By taking the dynamic support structure into account as well, we develop a DCS-based algorithm with both low complexity and enhanced performance in the following sections.
%As mentioned in the above, there has temporal correlation in the active user sets in the $T$ frames.
%This means that more accurate user activity detection and channel estimation performance in the current frame can be achieve with the assistance of the detected active user set in the previous frames. So that we can redesign the CS-based recovery algorithm to take the dynamic support structure into account as well.
The DCS problem can be solved by optimization-based methods using convex relaxation and the standard convex problem solver, e.g., Dynamic LASSO \cite{Angle_2009_ICDSP}. It can also be solved by the greed-based algorithms based on OMP and SP by employing the detected active user set in the previous frame as the prior information to assist the detection in the current frame \cite{Wang_2016_CL, Du_2017_JSAC}. These two approaches can outperform the traditional CS-based algorithms, but however, are unable to fully utilize the system statistics including the Markov chain-based user activity and the channel distributions.
To improve the performance over the optimization-based methods and the greedy-based algorithms, we propose to employ the Bayesian inference method in this work, whose details will be given in the next section.

%The sparse user activity pattern can be exploited to design

%In practice, the active users have a high probability to transmit data to the BS in adjacent frame in the mMTC service, which implies that the data transmission of one user may occupy several consecutive frames. So that there exits temporal correlation of the support of effective channel $\mathbf{x}$ in several continuous frames. Then the single frame static system is extended to the multiple continuous frames dynamic system, where the active user set changes slowly across the frames.
%
%Then the objective is to do joint user activity detection and channel estimation in $T$ continuous frames by recovering the $\mathbf{X} = [\mathbf{x}_{1},\mathbf{x}_{2},\dots,\mathbf{x}_T] \in \mathbb{C}^{N \times T}$ from the received signal $\mathbf{Y} = [\mathbf{y}_{1},\mathbf{y}_{2},\dots,\mathbf{y}_{T}] \in \mathbb{C}^{L \times T}$, where $\mathbf{x}_t$ is the effective channel vector in the $t$th frame. The received signal in the $T$ continuous frames can be represented as
%\begin{equation}\label{equ:rece_sig_T}
%    \mathbf{Y} = \mathbf{A}\mathbf{X} + \mathbf{W},
%\end{equation}
%where $\mathbf{Z} = [\mathbf{z}_1,\mathbf{z}_2,\dots,\mathbf{z}_t] \in \mathbb{C}^{L \times T}$ is the effective noise matrix in the $T$ continuous frames.

\section{Bayesian Inference}\label{sec:problem_formulation}

%The JUADCE problem in the multiple continuous frames can be regarded as an underdetermined linear inverse problem in the multiple measurement vector (MMV) form with dynamic support, which is also referred as \emph{dynamic compressed sensing}.
%Obviously, traditional CS-based algorithms for the MMV problem with common sparsity assumption are not suitable to our problem due to the dynamic support structure.

In this section, the probabilistic model of the temporal-correlated massive access system is first introduced, then we introduce the powerful Bayesian inference methods to solve the considered DCS problem.
%By accounting the system statistics, the Bayesian inference method can usually realize superior performance to the optimization-based algorithms and the greedy-based algorithms.
%However, the computational complexities of the optimization-based algorithms and the greedy-based algorithms are usually unaffordable in large-scale system.
%But the computation complexity of these algorithms is high when the number of users is large.

%However, the system statistical information including the channel distribution and the active probability of users are not exploited. To improve the recovery performance, we formulate the JUADCE problem in a probabilistic model and use Bayesian inference to solve it.
%To further enhance the performance of the recovery algorithm, we can additionally take advantage of the prior knowledge of the system statistics including the channel distribution and active probability of users. We also propose to simultaneously process the received signals in several consecutive frames, which can exploit the temporal correlation existing in the user activities of the adjacent frames. Thus, we first describe the temporally-correlated massive access system by a probabilistic model and introduce the powerful Bayesian inference methods for joint user activity detection and channel estimation.

%\subsection{Probabilistic Model}

First, we represent all the variables and their relationships from the probability perspective.
Since the block fading Rayleigh channel is assumed, the prior probability of the effective channel coefficient of user $n$ in the $t$th frame, $x_{n,t}$, can be characterized by an independent Bernoulli-Gaussian distribution as
\begin{equation}\label{equ:Ch_BG}
\abovedisplayskip=1.5pt
\belowdisplayskip=1.5pt
    p(x_{n,t}|\beta,\lambda_{n,t}) =  (1-\lambda_{n,t}) \delta(x_{n,t}) + \lambda_{n,t} \mathcal{CN}(x_{n,t};0,\beta), \forall n,t,
\end{equation}
where $\delta(\cdot)$ is the Dirac delta function. Note that the more complicated channel model can also be considered and we can approximate the channel distribution with the Gaussian mixture distribution \cite{Vila_2013_TSP}. The conditional probability $p(\mathbf{y}_{t}|\mathbf{x}_{t})$ is obtained from the AWGN channel as
\begin{equation}\label{equ:Cp_y}
\abovedisplayskip=1pt
\belowdisplayskip=1pt
    p(\mathbf{y}_{t}|\mathbf{x}_{t}) = \mathcal{CN}(\mathbf{y}_{t};\mathbf{A}\mathbf{x}_t,\sigma^2_w\mathbf{I}_L), \forall t.
\end{equation}
By combining the Markov-chain modeled user activities, the probabilistic signal model describing the linear system of (\ref{equ:rece_sig_T}) can be finally derived as
\begin{align}\label{equ:PSM}
    p(\mathbf{Y},\mathbf{X},\pmb{\Lambda}) &=  \prod_{t=1}^{T} \left( p(\mathbf{y}_t|\mathbf{x}_t) \prod_{n=1}^{N} p(x_{n,t}|\lambda_{n,t}) p(\lambda_{n,t}|\lambda_{n,t-1}) \right),
    %\vspace{0.5cm}
    %\Big(p(\mathbf{y}_1|\mathbf{x}_1) \prod_{n=1}^{N}p(x_{n,1}|\lambda_{n,1})p(\lambda_{n,1})\Big)
\end{align}
%{\setlength{\abovedisplayskip}{3pt}%
%	\setlength{\belowdisplayskip}{3pt}%
%	\setlength{\abovedisplayshortskip}{0pt}%
%	\setlength{\belowdisplayshortskip}{0pt}%
%\begin{align}
%    p(\mathbf{Y},\mathbf{X},\pmb{\Lambda}) &=  \prod_{t=1}^{T} \left( p(\mathbf{y}_t|\mathbf{x}_t) \prod_{n=1}^{N} p(x_{n,t}|\lambda_{n,t}) p(\lambda_{n,t}|\lambda_{n,t-1}) \right),
%\end{align}
%}
where it is noted that $p(\lambda_{n,1}|\lambda_{n,0}) = p(\lambda_{n,1}) = (1-p_a)(1-\lambda_{n,1})+p_a\lambda_{n,1}$ and $p(\lambda_{n,t}|\lambda_{n,t-1})$ is given as (\ref{equ:p_trans_lamb_tt-1}) in the Appendix.

%\subsection{Bayesian Inference}

Based on the probabilistic model (\ref{equ:PSM}), the optimal performance of user activity detection and channel estimation under the minimum mean square error (MMSE) principle can be realized by Bayesian inference.
%In specific, the marginal posterior distributions of the effective channel coefficients $\{x_{n,t}\}$ and the activity states $\{\lambda_{n,t}\}$ with the received signal $\mathbf{Y}$ are calculated by Bayes' rule. Then
%Compared with the greedy-based algorithms, the Bayesian inference framework iteratively updates the channel coefficients and the activity likelihood value, which makes use of the Markov chain model to reduce the estimation error.
%We define $\tilde{\mathbf{\Lambda}}$ to be the activity matrix set $\{\mathbf{\Lambda_t}\}_{t=1}^{T}$.
Specifically, we first follow the Bayes' rule to obtain the posterior joint probability of $\mathbf{X}$ and $\mathbf{\Lambda}$ as
\begin{align}\label{equ:Pp_X}
    p(\mathbf{X},\mathbf{\Lambda}|\mathbf{Y}) &= \frac{p(\mathbf{Y},\mathbf{X},\pmb{\Lambda})}{p(\mathbf{Y})}
    %%%\Bigg( \prod_{t=1}^{T} \Big( p(\mathbf{y}_t|\mathbf{x}_t)
    %\frac{1}{p(\mathbf{Y})} p(\mathbf{Y}|\mathbf{X}|\tilde{\mathbf{\Lambda}}) p(\tilde{\mathbf{\Lambda}}) \notag \\
    %\quad &= \frac{1}{p(\mathbf{Y})} \Bigg( \prod_{t=1}^{T} \Big( p(\mathbf{y}_t|\mathbf{x}_t) \notag \\
    %%\quad &\times
    %%%\prod_{n=1}^{N} p(x_{n,t}|\lambda_{n,t})p(\lambda_{n,t}|\lambda_{n,t-1}) \Big) \Bigg),
\end{align}
%where $p(\lambda_{n,1}|\lambda_{n,0}) = (1-p_a)(1-\lambda_{n,1})+p_a\lambda_{n,1}$, $p(x_{n,t}|\lambda_{n,t}) = (1-\lambda_{n,t})\delta(x_{n,t}) + \lambda_{n,t}\mathcal{CN}(x_{n,t};0,\beta)$, and
where the marginal probability of the received signal $\mathbf{Y}$ is derived by
\begin{align}\label{equ:p_Y}
    p(\mathbf{Y}) = \int_{\mathbf{X}} \int_{\mathbf{\mathbf{\Lambda}}}
    %\prod_{t=1}^{T} \left( p(\mathbf{y}_t|\mathbf{x}_t) \prod_{n=1}^{N} p(x_{n,t}|\lambda_{n,t}) p(\lambda_{n,t}|\lambda_{n,t-1}) dx_{n,t}d\lambda_{n,t} \right)
    p(\mathbf{Y},\mathbf{X},\pmb{\Lambda}) ~d\mathbf{X}d\mathbf{\mathbf{\Lambda}}.
\end{align}

Given the posterior marginal probability $p(x_{n,t}|\mathbf{Y}) = \int_{\mathbf{\Lambda}} d\mathbf{\Lambda} \int_{\mathbf{X}_{\setminus x_{n,t}}} p(\mathbf{X},\mathbf{\Lambda}|\mathbf{Y}) d\mathbf{X}_{\setminus x_{n,t}} $, the MMSE estimator for $x_{n,t}$ can be expressed as
\begin{equation}\label{equ:mmse_X}
\abovedisplayskip=1pt
\belowdisplayskip=1pt
    \widehat{x}_{n,t} = \int x_{n,t} \cdot p(x_{n,t}|\mathbf{Y}) ~dx_{n,t}, \forall n,t.
\end{equation}
%where $p(x_{n,t}|\mathbf{Y}) = \int_{\mathbf{\Lambda}} d\tilde{\mathbf{\Lambda}} \int_{\mathbf{X}_{\setminus x_{n,t}}} p(\mathbf{X},\mathbf{\Lambda}|\mathbf{Y}) d\mathbf{X}_{\setminus x_{n,t}} $ is the marginal posterior probability of $x_{n,t}$.
The estimator (\ref{equ:mmse_X}) achieves the minimum (Bayesian) MSE defined as
\begin{equation}\label{equ:mse_X}
    \text{MSE}(\mathbf{X}) = \frac{1}{NT}\mathbb{E}\left[||\widehat{\mathbf{X}} - \mathbf{X}||_F^2\right],
\end{equation}
where the expectation is computed over the joint distribution $p(\mathbf{X},\mathbf{\Lambda},\mathbf{Y})$ expressed in (\ref{equ:PSM}). Then the activity likelihood $\widehat{p}_{n,t}$ of user $n$ in frame $t$ can be obtained as
\begin{equation}\label{equ:rho_u}
\abovedisplayskip=1pt
\belowdisplayskip=1pt
    \widehat{p}_{n,t} = p(\lambda_{n,t}=1|\mathbf{Y}), \forall n, t,
\end{equation}
where $p(\lambda_{n,t}|\mathbf{Y}) = \int_{\mathbf{X}} d\mathbf{X} \int_{\mathbf{\Lambda}_{\setminus\lambda_{n,t}}} p(\mathbf{X},\mathbf{\Lambda}|\mathbf{Y}) d\mathbf{\Lambda}_{\setminus \lambda_{n,t}} $. By comparing (\ref{equ:rho_u}) with a predefined threshold, the user activities can be finally decided.

Due to the fact that the number of users in massive access is very large, both the MMSE estimator (\ref{equ:mmse_X}) and the user activity likelihood calculator (\ref{equ:rho_u}) involve very high-dimensional integrals and are thus intractable. In the next section, we will design a computationally efficient HyGAMP-based algorithm to approximately derive (\ref{equ:mmse_X}) and (\ref{equ:rho_u}) for our problem.

\section{The HyGAMP-DCS Algorithm}\label{sec:HyGAMP-DCS}

\begin{figure}[t]
  \centering
  \includegraphics[width=.5\textwidth]{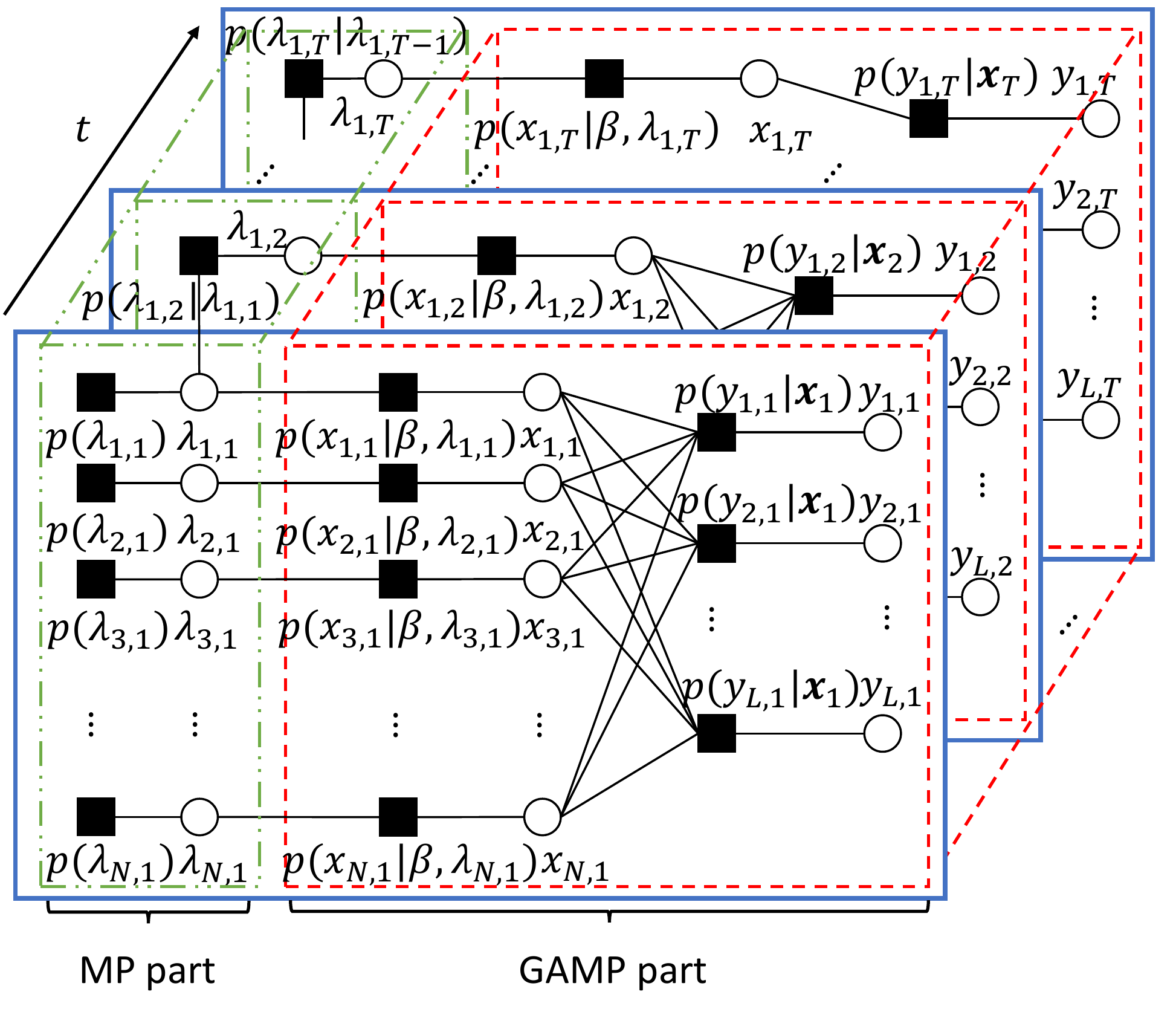}
  \vspace{-0.6cm}
  \caption{The factor graph representation of the joint posterior distribution $p(\mathbf{X},\mathbf{\Lambda}|\mathbf{Y})$.}\label{Fig:FG}
  \vspace{-0.5cm}
\end{figure}

%\textcolor[rgb]{0.00,0.07,1.00}{
%In this section, we propose a joint user activity detection and channel estimation based on the HyGAMP framework.
Before illustrating the algorithm design, we first give the factor graph representation of the joint marginal probability $p(\mathbf{Y}, \mathbf{X}, \pmb{\Lambda})$ with decomposition (\ref{equ:PSM}) as shown in Fig. \ref{Fig:FG}. Specifically, the black rectangle represents the factor node corresponding to the function $p(y_{l,t}|\mathbf{x}_{t})$, $p(x_{n,t}|\beta,\lambda_{n,t})$ or $p(\lambda_{n,t}|\lambda_{n,t-1})$, and the blank circle represents the variable node with the random variable $y_{l,t}$, $x_{n,t}$ or $\lambda_{n,t}$.
Intuitively, the factor graph suggests the usage of the MP algorithm \cite{Ksch_2001_TIT} to approximately obtain (\ref{equ:mmse_X}) and (\ref{equ:rho_u}). The standard MP algorithm iteratively updates the messages conveyed between the adjacent nodes and aggregates the messages arrived at the nodes $\{x_{n,t}\}$ and $\{\lambda_{n,t}\}$ to obtain their posterior probabilities, which avoids the extremely high-dimensional integrals in (\ref{equ:mmse_X}) and (\ref{equ:rho_u}).
However, the calculation of messages $\mu^i_{(n,t) \rightarrow (l,t)}(x_{n,t})$ and $\upsilon^i_{(n,t) \leftarrow (l,t)}(x_{n,t})$ in the dense bipartite graph including the nodes $\{x_{n,t}\}$ and $\{p(y_{l,t}|\mathbf{x}_{t})\}$ is still intractable since $L$ and $N$ is large in massive access. Fortunately, the HyGAMP framework \cite{Rangan_2017_TSP} can be leveraged to significantly simplify the implementation of the standard MP algorithm. Its key idea is to partition the edges in the factor graph into the strong and weak subsets. Then the messages passed on the weak edges are simplified by the AMP-style approximation, while the messages passed on the strong edge are still updated by standard MP principle.%}

%\textcolor[rgb]{0.00,0.07,1.00}{
By following the idea of the HyGAMP framework, we propose a HyGAMP-DCS algorithm specific to the considered DCS problem. Compared with the conventional HyGAMP algorithm in \cite{Rangan_2017_TSP}, the MP equations in HyGAMP-DCS are re-derived based on the statistical dependencies of the signals in our considered system.
The HyGAMP-DCS algorithm can be divided into two parts of GAMP and MP based on the message updating rule, where these two parts perform channel estimation and activity likelihood update, respectively. By exchanging the extrinsic messages between these two parts, the performance of user activity detection and channel estimation can both be improved. In the following, the details of the message passing equations in these two part are introduced.%}

\begin{table}[t]
  \centering
  \caption{Message Definition in the Factor Graph at the $i$th Iteration}
  \label{table:MD}
    \begin{tabular}{|l|l|}
    \hline
        $\mu^i_{(n,t) \rightarrow (l,t)}(x_{n,t})$ & message from $x_{n,t}$ to $p(y_{l,t}|\mathbf{x}_t)$ \\ \hline
        $\upsilon^i_{(n,t) \leftarrow (l,t)}(x_{n,t})$ & message from $p(y_{l,t}|\mathbf{x}_t)$ to $x_{n,t}$ \\ \hline
        $\mu^i_{(n,t) \rightarrow (n,t)}(x_{n,t})$ & message from $x_{n,t}$ to $p(x_{n,t}|\beta,\lambda_{n,t})$ \\  \hline
        $\upsilon^i_{(n,t) \leftarrow (n,t)}(x_{n,t})$ & message from $p(x_{n,t}|\beta,\lambda_{n,t})$ to $x_{n,t}$ \\  \hline
        $\mu^i_{(n,t) \rightarrow (n,t)}(\lambda_{n,t})$ & message from $\lambda_{n,t}$ to $p(x_{n,t}|\beta,\lambda_{n,t})$ \\  \hline
        $\upsilon^i_{(n,t) \leftarrow (n,t)}(\lambda_{n,t})$ & message from $p(x_{n,t}|\beta,\lambda_{n,t})$ to $\lambda_{n,t}$ \\  \hline
        $\xi^i_{(n,t) \rightarrow (n,t)}(\lambda_{n,t})$ & message from $p(\lambda_{n,t}|\lambda_{n,t-1})$ to $\lambda_{n,t}$ \\  \hline
        $\zeta^i_{(n,t) \leftarrow (n,t)}(\lambda_{n,t})$ & message from $\lambda_{n,t}$ to $p(\lambda_{n,t}|\lambda_{n,t-1})$ \\  \hline
        $\xi^i_{(n,t) \rightarrow (n,t+1)}(\lambda_{n,t})$ & message from $\lambda_{n,t}$ to $p(\lambda_{n,t+1}|\lambda_{n,t})$ \\  \hline
        $\zeta^i_{(n,t) \leftarrow (n,t+1)}(\lambda_{n,t})$ & message from $p(\lambda_{n,t+1}|\lambda_{n,t})$ to $\lambda_{n,t}$ \\
    \hline
	\end{tabular}
\vspace{-0.3cm}
\end{table}

\subsection{GAMP Part}

%This part estimates the effective channel matrix $\mathbf{X}$ based on the MMSE criterion with the extrinsic messages $\mu^i_{(n,t) \rightarrow (n,t)}(\lambda_{n,t}) = (1-\overrightarrow{p}_{n,t}(i))(1-\lambda_{n,t}) + \overrightarrow{p}_{n,t}(i)\lambda_{n,t}$ whose detailed derivation will be given as (\ref{equ:pr_nt}) in the MP part.
Following the HyGAMP framework, the GAMP part regards the edges between the nodes $\{x_{n,t}\}$ and $\{p(y_{l,t} |\mathbf{x}_t )\}$ as the weak edges thanks to their linearizable coupling. Then the messages $\mu^i_{(n,t) \rightarrow (l,t)}(x_{n,t})$ and $\upsilon^i_{(n,t) \leftarrow (l,t)}(x_{n,t})$ are simplified by using GAMP approximations based on the central limit theorem and Taylor expansion.

The basic GAMP estimation is given in lines 8-17 of Algorithm \ref{Alg:SP-HyGAMP-DCS}, which can incorporate arbitrary distribution on the input $\mathbf{X}$ and output $\mathbf{Y}$ \cite{Rangan_2011_ISIT}.
%The lines 9-18 of Algorithm \ref{Alg:SP-HyGAMP-DCS} give the basic GAMP estimation that can incorporate arbitrary distribution on the input $\mathbf{X}$ and output $\mathbf{Y}$ \cite{Rangan_2011_ISIT}.
Compared with the standard MP algorithm, the number of the update variables in GAMP estimation shrinks from $\mathcal{O}(LN)$ to $\mathcal{O}(L+N)$.
In lines 8-11, the GAMP estimation performs the update of the noiseless signal $\mathbf{Z} = \mathbf{A}\mathbf{X}$. The distribution of each element $z_{l,t}$ is obtained as $\mathcal{CN}(z_{l,t};\widehat{p}_{l,t}(i),\tau^p_{l,t}(i))$ by the messages from the variable nodes $\{x_{n,t}\}$, where $\widehat{p}_{l,t}(i)$ is the plug-in estimate of $z_{l,t}$ with variance $\tau^p_{l,t}(i)$ in the $i$th iteration.
%The update for $\widehat{z}_{l,t}^0$ and $\tau^z_{l,t}$ only depends on the output distributions where $\mathbf{Z} = \mathbf{A}\mathbf{X}$.
By accounting the AWGN channel between $\mathbf{Z}$ and $\mathbf{Y}$, the explicit expression of the MMSE estimator for $z_{l,t}$ and its variance in lines 10-11 are available as
\begin{align}
    \widehat{z}^0_{l,t}(i) &= \frac{y_{l,t}\tau^p_{l,t}(i)+\widehat{p}_{l,t}(i)\sigma^2_w}{\tau^p_{l,t}(i)+\sigma^2_w}, \label{equ:z0} \\
    \tau^z_{l,t}(i) &= \frac{\tau^p_{l,t}(i)\sigma^2_w}{\tau^p_{l,t}(i)+\sigma^2_w}. \label{equ:tau_w}
\end{align}
%\textcolor[rgb]{0.00,0.07,1.00}{
After updating the noiseless signal $z_{n,t}$, the mean and the variance of each element in the residual signal that removes the estimated $\mathbf{X}$ from $\mathbf{Y}$ are given in lines 12-13. Then lines 14-15 give the update of the plug-in estimate $\widehat{r}_{n,t}$ of the true signal $x_{n,t}$, which is modeled as $\widehat{r}_{n,t}(i) = x_{n,t} + n_{n,t}(i)$ with $n_{n,t}(i) \sim \mathcal{CN}(0,\tau^r_{n,t}(i) )$.
%} %$p(\widehat{r}_{n,t}|x_{n,t}) = \mathcal{CN}(\widehat{r}_{n,t};x_{n,t},\tau^r_{n,t})$.
To obtain the MMSE estimator of $\mathbf{X}$, we can give the approximate posterior marginal distribution of $x_{n,t}$ as
%\textcolor[rgb]{0.00,0.07,1.00}{
\begin{align}\label{equ:Pp_x}
    p^i(x_{n,t}|\mathbf{Y}) = \frac{p(\widehat{r}_{n,t}(i)) p(\widehat{r}_{n,t}(i)|x_{n,t})}{p(\widehat{r}_{n,t}(i))} = \frac{p^i_{X}(x_{n,t}) \mathcal{CN}(x_{n,t};\widehat{r}_{n,t}(i),\tau_{n,t}^r(i))}{\int p^i_{X}(x_{n,t})\mathcal{CN}(x_{n,t};\widehat{r}_{n,t}(i),\tau_{n,t}^r(i))dx_{n,t}},
\end{align}
where $p^i_{X}(x_{n,t}) = (1-\overrightarrow{p}_{n,t}(i))\delta(x_{n,t}) + \overrightarrow{p}_{n,t}(i)\mathcal{CN}(x_{n,t};0,\beta)$ based on the extrinsic message $\mu^i_{(n,t)\rightarrow(n,t)}(\lambda_{n,t})=(1-\overrightarrow{p}_{n,t}(i))(1-\lambda_{n,t}) + \overrightarrow{p}_{n,t}(i)\lambda_{n,t}$. The detailed derivation of $\overrightarrow{p}_{n,t}(i)$ will be given as (\ref{equ:pr_nt}) in the MP part.
%\textcolor[rgb]{0.00,0.07,1.00}{
By simplifying (\ref{equ:Pp_x}), the approximate posterior marginal distribution $p^i(x_{n,t}|\mathbf{Y})$ can be written in the form of a spike and slab probability as
\begin{small}
\begin{align}\label{equ:Pp_x_ssp}
    &p^i(x_{n,t}|\mathbf{Y}) \notag \\
     &= \frac{(1-\overrightarrow{p}_{n,t}(i))\delta(x_{n,t})\mathcal{CN}(x_{n,t};\widehat{r}_{n,t}(i),\tau_{n,t}^r(i)) + \overrightarrow{p}_{n,t}(i)\mathcal{CN}(x_{n,t};0,\beta)\mathcal{CN}(x_{n,t};\widehat{r}_{n,t}(i),\tau_{n,t}^r(i))}{\int (1-\overrightarrow{p}_{n,t}(i))\delta(x_{n,t})\mathcal{CN}(x_{n,t};\widehat{r}_{n,t}(i),\tau_{n,t}^r(i)) + \overrightarrow{p}_{n,t}(i)\mathcal{CN}(x_{n,t};0,\beta)\mathcal{CN}(x_{n,t};\widehat{r}_{n,t}(i),\tau_{n,t}^r(i)) dx_{n,t}} \notag \\
    &= \frac{(1-\overrightarrow{p}_{n,t}(i))\mathcal{CN}(0;\widehat{r}_{n,t}(i),\tau_{n,t}^r(i))\delta(x_{n,t}) + \overrightarrow{p}_{n,t}(i)\mathcal{CN}(0;\widehat{r}_{n,t}(i),\beta+\tau^r_{n,t}(i))\mathcal{CN}(x_{n,t};\gamma_{n,t}(i),\tau^{\gamma}_{n,t}(i))}{(1-\overrightarrow{p}_{n,t}(i)) \mathcal{CN}(0;\widehat{r}_{n,t}(i),\tau_{n,t}^r(i)) + \overrightarrow{p}_{n,t}(i)\mathcal{CN}(0;\widehat{r}_{n,t}(i),\beta+\tau^r_{n,t}(i))} \notag \\
    &= (1-\varpi_{n,t}(i))\delta(x_{n,t}) + \varpi_{n,t}(i)\mathcal{CN}(x_{n,t};\gamma_{n,t}(i),\tau^{\gamma}_{n,t}(i)),
\end{align}
\end{small}where
\begin{align}
  \varpi_{n,t}(i) &=\left[1+\frac{1-\overrightarrow{p}_{n,t}(i)}{\overrightarrow{p}_{n,t}(i)}\frac{\beta+\tau_{n,t}^r(i)}{\tau_{n,t}^r(i)} \exp\Bigg(-\frac{|\widehat{r}_{n,t}(i)|^2}{\tau_{n,t}^r(i)(1+\frac{\tau_{n,t}^r(i)}{\beta})}\Bigg)\right]^{-1}, \label{equ:varpi}\\
  %\left( 1+\frac{1-\rho_{n,t}}{\rho_{n,t}}\frac{\beta_{n,t}+\tau_{n,t}^r}{\tau_{n,t}^r}\text{exp}\Big(-\frac{|\widehat{r}_{n,t}|^2}{\tau_{n,t}^r(1+\frac{\tau_{n,t}^r}{\beta_{n,t}})}\Big) \right)^{-1}, \label{equ:varpi}\\
  \gamma_{n,t}(i) &= \frac{\beta}{\beta+\tau_{n,t}^r(i)}\widehat{r}_{n,t}(i), \label{equ:gamma}\\
  \tau_{n,t}^{\gamma}(i) &= \frac{\beta\tau^r_{n,t}(i)}{\beta+\tau^r_{n,t}(i)}. \label{equ:tau_gamma}
\end{align}
Thus, we can obtain the explicit expression of the posterior mean and variance of $x_{n,m}$ in lines 16-17 as
\begin{align}
  \widehat{x}_{n,t}(i+1) &= \mathbb{E}[ x_{n,t} | \widehat{r}_{n,t}(i), \tau^r_{n,t}(i), \overrightarrow{p}_{n,t}(i) ] =  \varpi_{n,t}(i)\gamma_{n,t}(i), \label{equ:E_x}\\
  \tau^x_{n,t}(i+1) &= \mathbb{V}[ x_{n,t} | \widehat{r}_{n,t}(i), \tau^r_{n,t}(i), \overrightarrow{p}_{n,t}(i) ] = \varpi_{n,t}(i)\left((1-\varpi_{n,t}(i))|\gamma_{n,t}(i)|^2+\tau^{\gamma}_{n,t}(i)\right). \label{equ:V_x}
\end{align}
%\textcolor[rgb]{0.00,0.07,1.00}{
Since $\overrightarrow{p}_{n,t}(i)$ is updated by the MP part at each iteration, the MMSE estimator (\ref{equ:E_x}) also changes at each iteration. However, the MMSE estimator on $x_{n,t}$ keeps constant at each iteration in the AMP-based algorithms proposed in \cite{Jiang_2021_TWC, Wang_2021_ISIT} for temporally-correlated massive access.%}
%Compared with the AMP-based algorithms proposed in \cite{Jiang_2021_TWC, Wang_2021_arxiv} which have constant MMSE estimator on $x_{n,t}$ in each iteration, the MMSE estimator (\ref{equ:E_x}) in HyGAMP-DCS is different in each iteration since $\widehat{\rho}_{n,t}(i)$ is also updated in each iteration.

\subsection{MP Part}

\begin{algorithm}[t] \footnotesize
\caption{HyGAMP-DCS for joint user activity detection and channel estimation}\label{Alg:SP-HyGAMP-DCS}
\begin{algorithmic}[1]
\REQUIRE Pilot matrix $\mathbf{A}$, received signals $\mathbf{Y}$, large-scale fading coefficients $\mathbf{\beta}$, active probability $p_a$, transition probability matrix $\mathbf{P}$, noise variance $\sigma^2_w$, tolerance $\varepsilon$, and maximum iteration $I_{max}$
\ENSURE MMSE estimate $\widehat{\mathbf{X}}$
\STATE \textbf{Initialize}
\STATE $i \leftarrow 1$
\STATE $\forall n,t:~ \widehat{x}_{n,t}(i) = 0$, $\tau^x_{n,t}(i) = \mathbb{V}[x_{n,t}] = p_a \beta$, $\overrightarrow{p}_{n,t}(i) = p_a$.
\STATE $\forall l,t:~ \widehat{s}_{l,t}(i-1) = 0$.
\STATE $\forall n:~ \overrightarrow{q}_{n,1}(i) = p_a$.
\REPEAT
    \STATE \{Basic GAMP estimation\}
    \STATE $\forall l,t: ~\tau^p_{l,t}(i) =  \sum_{n=1}^{N} |a_{l,n}|^2 \tau^x_{n,t}(i)$.
    \STATE $\forall l,t: ~\widehat{p}_{l,t}(i) = \sum_{n=1}^{N} a_{l,n}\widehat{x}_{n,t}(i) - \tau^p_{l,t}(i)\widehat{s}_{l,t}(i-1)$.
    \STATE $\forall l,t: ~\tau^z_{l,t}(i) = \mathbb{V}[ z_{l,t} | \widehat{p}_{l,t}(i), \tau^p_{l,t}(i), y_{l,t}, \sigma^2_w ]$.
    \STATE $\forall l,t: ~\widehat{z}^0_{l,t}(i) = \mathbb{E}[ z_{l,t} | \widehat{p}_{l,t}(i), \tau^p_{l,t}(i), y_{l,t}, \sigma^2_w ]$.
    \STATE $\forall l,t: ~\tau^s_{l,t}(i) = (1-\tau^z_{l,t}(i)/\tau^p_{l,t}(i))/\tau^p_{l,t}(i)$.
    \STATE $\forall l,t: ~\widehat{s}_{l,t}(i) = (\widehat{z}^0_{l,t}(i)-\widehat{p}_{l,t}(i))/\tau^p_{l,t}(i)$.
    \STATE $\forall n,t: ~\tau^r_{n,t}(i) = 1/\left(\sum_{l=1}^{L}|a_{l,n}|^2\tau^s_{l,n}(i)\right)$.
    \STATE $\forall n,t: ~\widehat{r}_{n,t}(i) = \widehat{x}_{n,t}(i) + \tau^r_{n,t}(i)\sum_{l=1}^{L}a^{*}_{l,t}\widehat{s}_{l,t}(i)$.
    \STATE $\forall n,t: ~\tau^x_{n,t}(i+1) = \mathbb{V}[ x_{n,t} | \widehat{r}_{n,t}(i), \tau^r_{n,t}(i), \overrightarrow{p}_{n,t}(i) ]$.
    \STATE $\forall n,t: ~\widehat{x}_{n,t}(i+1) = \mathbb{E}[ x_{n,t} | \widehat{r}_{n,t}(i), \tau^r_{n,t}(i), \overrightarrow{p}_{n,t}(i) ]$.
    \STATE \{Activity likelihood update\}
    \STATE $\forall n,t: ~\overleftarrow{p}_{n,t}(i)$ is updated in (\ref{equ:pl_nt}).
    \STATE $\forall n,t: ~\overrightarrow{q}_{n,t}(i)$ is updated in (\ref{equ:qr_nt}) from frame 1 to frame $T$.
    \STATE $\forall n,t: ~\overleftarrow{q}_{n,t}(i)$ is updated in (\ref{equ:ql_nt}) from frame $(T-1)$ to frame 1.
    \STATE $\forall n,t: ~\overrightarrow{p}_{n,t}(i+1)$ is updated in (\ref{equ:pr_nt}).
    %\STATE $\forall n,t: ~\widehat{\rho}_{n,t}(i+1) = \overrightarrow{p}_{n,t}(i)$.
    \STATE $i \leftarrow i+1$
\UNTIL{$i > I_{max}$ or $\frac{||\widehat{\mathbf{X}}(i+1)-\widehat{\mathbf{X}}(i)||_F^2}{||\widehat{\mathbf{X}}(i)||_F^2} \le \varepsilon$}.
\end{algorithmic}
\end{algorithm}

By regarding all the edges in the MP part as strong edges, we utilize the standard MP algorithm to update the activity probability of each user with the extrinsic message $\upsilon^i_{(n,t) \leftarrow (n,t)}(\lambda_{n,t})$ from the GAMP part.
%\textcolor[rgb]{0.00,0.07,1.00}{Compared with S-AMP \cite{Jiang_2021_TWC} only performing forward message passing from $\lambda_{n,t-1}$ to $\lambda_{n,t}$, the MP part in HyGAMP-DCS updates both the forward message from $\lambda_{n,t-1}$ to $\lambda_{n,t}$ and the backward message from $\lambda_{n,t+1}$ to $\lambda_{n,t}$. The details of the forward and backward messages are given in the following.}

First, the extrinsic message $\upsilon^i_{(n,t) \leftarrow (n,t)}(\lambda_{n,t})$ is obtained as
\begin{align}\label{equ:Ii}
    \upsilon^i_{(n,t) \leftarrow (n,t)}(\lambda_{n,t}) = & \int p(x_{n,t}|\beta,\lambda_{n,t}) \cdot \upsilon^i_{(n,t) \leftarrow (n,t)}(x_{n,t}) dx_{n,t} \notag \\
%    \upsilon_{(n,t) \leftarrow (n,t)}(\lambda_{n,t}) \propto& \int p(x_{n,t}|\beta_{n,t},\lambda_{n,t}) \notag \\
%    \quad &\times \upsilon_{(n,t) \leftarrow (n,t)}(x_{n,t}) dx_{n,t}, \notag \\
    \quad = &~ (1-\overleftarrow{p}_{n,t}(i))(1-\lambda_{n,t}) + \overleftarrow{p}_{n,t}(i)\lambda_{n,t},
\end{align}
where $\upsilon^i_{(n,t) \leftarrow (n,t)}(x_{n,t}) = \mathcal{CN}(x_{n,t};\widehat{r}_{n,t}(i),\tau^r_{n,t}(i))$ and $\overleftarrow{p}_{n,t}(i)$ can be expressed by
%\textcolor[rgb]{0.00,0.07,1.00}{
\begin{align}\label{equ:pl_nt}
    \overleftarrow{p}_{n,t}(i) &= \frac{p(\widehat{r}_{n,t}(i)|\lambda_{n,t}=1)}{p(\widehat{r}_{n,t}(i)|\lambda_{n,t}=0)+p(\widehat{r}_{n,t}(i)|\lambda_{n,t}=1)} \notag \\
    \quad &=\frac{\mathcal{CN}(\widehat{r}_{n,t};0,\tau^r_{n,t}(i)+\beta)}{\mathcal{CN}(\widehat{r}_{n,t};0,\tau^r_{n,t}(i))+\mathcal{CN}(\widehat{r}_{n,t};0,\tau^r_{n,t}(i)+\beta)}.
\end{align}
With (\ref{equ:Ii}), we can implement the statistical dependencies across the frames based on the Markov chain of the activity variables $\{\lambda_{n,t}\}$. We also define $\overleftarrow{q}_{n,t}(i)$ and $\overrightarrow{q}_{n,t}(i)$ as the activity likelihoods of user $n$ contained in the messages $\zeta^i_{(n,t) \leftarrow (n,t+1)}(\lambda_{n,t})$ and $\xi^i_{(n,t) \rightarrow (n,t)}(\lambda_{n,t})$, respectively.
%The messages $\xi_{(n,t-1) \rightarrow (n,t)}(\lambda_{n,t-1})$ is calculated by the product of two Bernoulli distributions
%\begin{align}\label{equ:m_xi_nt-1_nt}
%  \xi_{(n,t-1) \rightarrow (n,t)}(\lambda_{n,t-1}) \propto &~ (1-q_{n,t-1})(1-p_{n,t-1})(1-\lambda_{n,t-1}) \notag \\
%    \quad &~ + q_{n,t-1}p_{n,t-1}\lambda_{n,t-1},
%\end{align}
%The message update equations in the MP part are given in the following
%\begin{align}
%    \xi_{(n,t) \rightarrow (n,t)}(\lambda_{n,t}) \propto &~ \int \xi_{(n,t-1) \rightarrow (n,t)}(\lambda_{n,t-1}) \notag \\
%    \quad &~ \times p(\lambda_{n,t}|\lambda_{n,t-1}) d\lambda_{n,t-1}, \\
%    \zeta_{(n,t) \leftarrow (n,t)}(\lambda_{n,t}) \propto &~ \upsilon_{(n,t) \leftarrow (n,t)}(\lambda_{n,t}) \notag \\
%    \quad &~ \times \zeta_{(n,t) \leftarrow (n,t+1)}(\lambda_{n,t}), \\
%    \xi_{(n,t-1) \rightarrow (n,t)}(\lambda_{n,t-1}) \propto &~ (1-\overrightarrow{q}_{n,t-1})(1-\overleftarrow{p}_{n,t-1})(1-\lambda_{n,t-1}) \notag \\
%    \quad &~ + \overrightarrow{q}_{n,t-1}\overleftarrow{p}_{n,t-1}\lambda_{n,t-1}, \\
%    \zeta_{(n,t-1) \leftarrow (n,t)}(\lambda_{n,t-1}) \propto &~ \int \zeta_{(n,t) \leftarrow (n,t)}(\lambda_{n,t}) \notag \\
%    \quad &~ \times p(\lambda_{n,t}|\lambda_{n,t-1}) d\lambda_{n,t},
%\end{align}
%Due to the page limit, some details of derivation of the updated messages are discarded in the following.
The message $\xi^i_{(n,t) \rightarrow (n,t+1)}(\lambda_{n,t})$ for $\forall t = 1,\dots,T-1$ is expressed by the product of two Bernoulli distributions as
%\textcolor[rgb]{0.00,0.07,1.00}{
\begin{align}\label{equ:xi_nt-1_nt}
    \xi^i_{(n,t) \rightarrow (n,t+1)}(\lambda_{n,t}) \propto &~ \xi^i_{(n,t) \rightarrow (n,t)}(\lambda_{n,t}) \cdot \upsilon^i_{(n,t) \leftarrow (n,t)}(\lambda_{n,t}) \notag \\
    \quad  \propto &~ (1-\overrightarrow{q}_{n,t}(i))(1-\overleftarrow{p}_{n,t}(i))(1-\lambda_{n,t})
     + \overrightarrow{q}_{n,t}(i)\overleftarrow{p}_{n,t}(i)\lambda_{n,t}.
%    \xi_{(n,t) \rightarrow (n,t+1)}(\lambda_{n,t}) \propto &~ (1-\overrightarrow{q}_{n,t})(1-\overleftarrow{p}_{n,t})(1-\lambda_{n,t}) \notag \\
%    \quad &~ + \overrightarrow{q}_{n,t}\overleftarrow{p}_{n,t}\lambda_{n,t}.
\end{align}
Then we update the forward message $\xi^i_{(n,t) \rightarrow (n,t)}(\lambda_{n,t})$ that represents the useful information conveyed from $\lambda_{n,t-1}$ to $\lambda_{n,t}$. For $t=1$, we always $\xi^i_{(n,t) \rightarrow (n,t)}(\lambda_{n,t}) = (1-p_a)(1-\lambda_{n,t}) + p_a\lambda_{n,t}$, i.e., $\overrightarrow{q}_{n,1}(i) = p_a, \forall n$, since $p(\lambda_{n,1})$ is only connected with $\lambda_{n,1}$ in the factor graph. For $t\ge2$, the forward message $\xi^i_{(n,t) \rightarrow (n,t)}(\lambda_{n,t})$ can be obtained as
\begin{align}\label{equ:xi_nt_nt}
    \xi^i_{(n,t) \rightarrow (n,t)}(\lambda_{n,t}) = & \int p(\lambda_{n,t}|\lambda_{n,t-1}) \cdot \xi^i_{(n,t-1) \rightarrow (n,t)}(\lambda_{n,t-1}) d\lambda_{n,t-1} \notag
    %\quad = &~ (1-\overrightarrow{q}_{n,t}(i))(1-\lambda_{n,t}) + \overrightarrow{q}_{n,t}(i)\lambda_{n,t},
%    \xi_{(n,t+1) \rightarrow (n,t+1)}(\lambda_{n,t+1}) = &~ (1-\overrightarrow{q}_{n,t+1})(1-\lambda_{n,t+1}) \notag \\
%    \quad &~ + \overrightarrow{q}_{n,t+1}\lambda_{n,t+1},
\end{align}
where
\begin{equation}\label{equ:qr_nt}
    \overrightarrow{q}_{n,t}(i) = \frac{p_{01}(1-\overleftarrow{p}_{n,t-1}(i))(1-\overrightarrow{q}_{n,t-1}(i))+p_{11}\overleftarrow{p}_{n,t-1}(i)\overrightarrow{q}_{n,t-1}(i)}
    {(1-\overleftarrow{p}_{n,t-1}(i))(1-\overrightarrow{q}_{n,t-1}(i))+\overleftarrow{p}_{n,t-1}(i)\overrightarrow{q}_{n,t-1}(i)}.
\end{equation}
Note that the activity likelihoods $\{\overrightarrow{q}_{n,t}(i)\}$ are updated sequentially from frame $1$ to frame $T$.%}

During the forward message passing, the backward message passing can be performed in parallel to convey the useful information from $\lambda_{n,t+1}$ to $\lambda_{n,t}$ for $t = 1, \dots, T-1$.
Due to the symmetry, the update formulas for $\overleftarrow{q}_{n,t}(i)$ in the backward message $\zeta^i_{(n,t)\leftarrow(n,t+1)}(\lambda_{n,t})$ is similar to that of $\overrightarrow{q}_{n,t}(i)$, except that %\textcolor[rgb]{0.00,0.07,1.00}{
$\{\overleftarrow{q}_{n,t}(i)\}$ are updated sequentially from frame $(T-1)$ to frame $1$. For $t \le T-1$, the activity likelihood $\overleftarrow{q}_{n,t}(i)$
%in the backward message $\zeta^i_{(n,t)\leftarrow(n,t+1)}(\lambda_{n,t})$
can be obtained as%}
%\begin{equation}\label{equ:zetal_t+1t}
%\abovedisplayskip=1pt
%\belowdisplayskip=1pt
%    \zeta^i_{(n,t)\leftarrow(n,t+1)}(\lambda_{n,t}) = (1-\overleftarrow{q}_{n,t}(i))(1-\lambda_{n,t}) + \overleftarrow{q}_{n,t}(i)\lambda_{n,t},
%\end{equation}
%where
\begin{equation}\label{equ:ql_nt}
    \overleftarrow{q}_{n,t}(i) = \frac{p_{10}(1-\overleftarrow{p}_{n,t+1}(i))(1-\overleftarrow{q}_{n,t+1}(i)) + p_{11}\overleftarrow{p}_{n,t+1}(i)\overleftarrow{q}_{n,t+1}(i)}{(p_{00}+p_{10})(1-\overleftarrow{p}_{n,t+1}(i))(1-\overleftarrow{q}_{n,t+1}(i)) + (p_{11}+p_{01})\overleftarrow{p}_{n,t+1}(i)\overleftarrow{q}_{n,t+1}(i)}.
\end{equation}
%For the special case $t=T$, the probability $\overleftarrow{q}_{n,T}$ can be obtained from (\ref{equ:ql_nt}) by setting $\overleftarrow{q}_{n,T+1} = 0.5$.
%\textcolor[rgb]{0.00,0.07,1.00}{
It is noted that the proposed algorithms with block-by-block detection performs bidirectional message propagation, while the algorithms in \cite{Jiang_2021_TWC, Wang_2021_ISIT} only have forward message propagation. Therefore, by utilizing the useful information from both $\lambda_{n,t-1}$ and $\lambda_{n,t+1}$, HyGAMP-DCS can further exploit the temporal correlation to provide a more precise estimation of the activity $\lambda_{n,t}$, which then also contributes to enhanced channel estimation performance.

With both the forward message and backward message in the above, the extrinsic message $\mu^i_{(n,t)\rightarrow(n,t)}(\lambda_{n,t})$ to refine the channel estimation in the next iteration is subsequently given by $\mu^i_{(n,t)\rightarrow(n,t)}(\lambda_{n,t}) \propto \xi^i_{(n,t) \rightarrow (n,t)}(\lambda_{n,t}) \cdot \zeta^i_{(n,t)\leftarrow(n,t+1)}(\lambda_{n,t})$ for $t = 1,\dots,T-1$.
Then the update rule of $\overrightarrow{p}_{n,t}(i)$ in message $\mu^i_{(n,t)\rightarrow(n,t)}(\lambda_{n,t})$ for $t = 1,\dots,T-1$ is obtained as
\begin{align}\label{equ:pr_nt}
    \overrightarrow{p}_{n,t}(i+1) = \frac{\overleftarrow{q}_{n,t}(i)\overrightarrow{q}_{n,t}(i)}{(1-\overleftarrow{q}_{n,t}(i))(1-\overrightarrow{q}_{n,t}(i))+\overleftarrow{q}_{n,t}(i)\overrightarrow{q}_{n,t}(i)}.
\end{align}
%\textcolor[rgb]{0.00,0.07,1.00}{
However, for $t=T$, the update rule is given by $\mu^i_{(n,t)\rightarrow(n,t)}(\lambda_{n,t}) = \xi^i_{(n,t) \rightarrow (n,t)}(\lambda_{n,t})$ and $\overrightarrow{p}_{n,T}(i+1) = \overrightarrow{q}_{n,T}(i)$, since the variable node $\lambda_{n,T}$ is connected to only one factor node $p(\lambda_{n,T}|\lambda_{n,T-1})$ in the MP part.
The whole procedure of the proposed HyGAMP-DCS algorithm is outlined in Algorithm \ref{Alg:SP-HyGAMP-DCS}.
%In specific, the line 21-25 conclude the message update for user activity in the MP part. The initialization of the algorithm is also given in line 2-7.
In this work, we only consider the algorithm design single-antenna scenario, while the extension of the algorithm for the multiple-antenna scenario is straightforward.

\subsection{Discussion}

In this subsection, the impact of the number of joint detection frames $T$ on the estimation performance and the detection delay in the proposed algorithm is discussed.
Due to the block-by-block detection, the estimation in the 1st frame and the $T$th frame only benefits from single-sided message, while the estimation in the $t$th frame ($2 \le t \le T-1$) can utilize the double-sided messages. The ratio of the number of frames in which the estimations can combine the double-sided messages to the number of joint detection frames, i.e., $\frac{T-2}{T}$, is enlarged when $T$ increases. By increasing the number of joint detection frames $T$, the overall user activity detection and channel estimation performance in the $T$ consecutive frames is improved. When $T$ is large, the ratio $\frac{T-2}{T}$ approaches $1$ and then the proposed algorithm may achieve its best performance.
%Compared with the DCS-based algorithms in \cite{Wang_2016_CL,Du_2017_JSAC,Jiang_2021_TWC,Wang_2021_ISIT}, the proposed algorithm achieves a tradeoff between the performance and the detection delay.
%and finally approach the saturated performance when $T$ is large enough.
However, the block-by-block detection also leads to detection delay, and the average detection delay can be obtained as $\frac{T-1}{2}$ (frame).
%From the user activity model in Section \ref{subsec:UAM}, the user activity in the $t$th frame is only statistically correlated with those in the $(t-1)$th and $(t+1)$th frame, but independent to those in the other frames. Thus, the estimation in the $t$th frame mainly benefits from the estimation results from the $(t-1)$th frame and the $(t+1)$th frame, while the estimation results in the farther frames provide much less performance gain.
%When the number of joint detection frames increases, the number of frames in which the estimations can combine the double-sided messages to refine the detection and estimation also increases. Thus, the overall performance is improved.}
%Due to the block-by-block detection, the proposed algorithm achieves a tradeoff between the performance and the detection delay. Though increasing $T$ can improve the performance, it also brings about larger detection delay.
%However, the proposed method brings about detection delay compared with \cite{Jiang_2021_TWC, Wang_2021_ISIT}, \textcolor[rgb]{0.00,0.07,1.00}{meaning that the proposed algorithm achieves a tradeoff between the performance and the detection delay.}
From the simulation results in Section \ref{sec:simulation}, we find that HyGAMP-DCS can achieve significant performance improvement with a moderate value of $T$ (e.g., $T=5$).

%\begin{remark}
%    The proposed HyGAMP-DCS algorithm has similar structure to the DCS-AMP algorithm proposed in \cite{Ziniel_2013_TSP}. However, the DCS-AMP in each iteration runs a complete AMP algorithm where the variables are usually updated for tens of times, while the variables in the GAMP part of the HyGAMP-DCS algorithm is only updated once. Note that the running time cost of the DCS-AMP and HyGAMP-DCS mainly results from calculation in the AMP estimation. So that the whole complexity of DCS-AMP are higher than that of HyGAMP, though the DCS-AMP may only need several iterations.
%\end{remark}

\section{Hyperparameter Learning via EM Algorithm}\label{sec:EM}

The HyGAMP-DCS algorithm requires the prior knowledge of the system statistics characterized by the hyperparameter set $\pmb{\vartheta}=\{p_a,\beta,p_{10},\sigma^2_w\}$, which may not be estimated accurately in advance.
%On the other hand, these statistical parameters in $\pmb{\vartheta}$ may be tuned based on the machine learning techniques to better match the currently received signals at the BS, which thus improves the performance.
In this section, we propose to integrate the expectation maximization algorithm \cite{Moon_1996_SPM} with the proposed HyGAMP-DCS algorithm, where the statistical parameters are learned from the received signals in the jointly detected $T$ frames during the estimation procedure.
% of the detection and estimation on $\mathbf{X}$ and $\pmb{\Lambda}$.
%\textcolor[rgb]{0.00,0.07,1.00}{
%Compared with the work \cite{Vila_2013_TSP} that concentrates on the CS problem with single measurement vector (SMV), our proposed algorithm solves the DCS-SMV problem and additionally learns the transition probability matrix $\mathbf{P}$ of the Markov chain. Moreover, the parameters of the channel statistics are learned from the received signals $\mathbf{Y}$ in multiple consecutive frames, which results in a more accurate hyperparameter learning result.%}

The EM algorithm aims to find the optimal parameter setting that maximizes the likelihood function $p(\mathbf{Y},\mathbf{X}|\pmb{\vartheta})$, while the optimization problem is usually non-convex and has intractable computational complexity. Instead, the EM algorithm maximizes a lower bound of the likelihood function $p(\mathbf{Y},\mathbf{X}|\pmb{\vartheta})$ in each iteration, then $p(\mathbf{Y},\mathbf{X}|\pmb{\vartheta})$ is guaranteed to be increased to a local maximal point. In specific, two steps of the E-step and M-step are alternatively performed until convergence, which can be written as
\begin{align}\label{equ:EM}
     \mathcal{L}^i(\pmb{\vartheta}) &= \mathbb{E}_{p^i(\mathbf{X})}[\ln p(\mathbf{X},\mathbf{Y}|\pmb{\vartheta})], \\
     %p^i(\mathbf{X}) &= p(\mathbf{X}|\mathbf{Y},\pmb{\vartheta}(i)), \\
     \pmb{\vartheta}^{i+1} &= \arg\max_{\pmb{\vartheta}} \mathcal{L}^i(\pmb{\vartheta}), \label{equ:M_step}
\end{align}
where $p^i(\mathbf{X}) = p(\mathbf{X}|\mathbf{Y},\pmb{\vartheta}^i)$ is the estimated posterior marginal distribution of $\mathbf{X}$ based on the updated statistical parameters $\pmb{\vartheta}^i$ and $\mathbb{E}_{p^{i}(\mathbf{x})}[\cdot]$ denotes the expectation over the posterior distribution $p^{i}(\mathbf{X})$. Since the number of users $N$ is very large, the expectation calculation may bring about unaffordable computational cost. Fortunately, the HyGAMP framework enables us to approximate the posterior probability by $p^i(\mathbf{X})=\prod_{t=1}^{T}\prod_{n=1}^{N}p(x_{n,t}|\mathbf{Y},\pmb{\vartheta}^i)$ in the large-scale systems, which leads to a computationally efficient EM estimation procedure. After obtaining the expectation, we leverage the EM algorithm in the incremental version \cite{Neal_1998} to simplify the joint optimization problem (\ref{equ:M_step}), where the hyperparameters are optimized in the coordinate-wise manner. So that only one hyperparameter $\vartheta \in \pmb{\vartheta}$ is updated at a time while others all keep fixed. The simplified optimization problem can be written as
%Then the simplified problem in the M-step can be represented as
\begin{align}\label{equ:Simp_M_step}
%\abovedisplayskip=1pt
%\belowdisplayskip=1pt
    %\vartheta &= \arg\max_{\vartheta} \mathcal{L}(\pmb{\vartheta}).
    \vartheta^{i+1} &= \arg\max_{\vartheta}\mathcal{L}^i(\pmb{\vartheta}^i_{\setminus \vartheta}, \vartheta).
\end{align}
The solution of the problem (\ref{equ:Simp_M_step}) can be easily obtained by setting the derivative of $\mathcal{L}^i(\pmb{\vartheta}^i_{\setminus \vartheta}, \vartheta)$ with respect to $\vartheta$ to be zero. The hyperparameters in the $i$th iteration are updated by
\begin{align}
    \sigma^2_w(i+1) =&~ \frac{1}{LT}\sum_{t=1}^{T}\sum_{l=1}^{L} \Big[|y_{l,t}-\widehat{z}^0_{l,t}(i)|^2+\tau_{l,t}^z(i)\Big], \label{equ:EM_tau_w} \\
    \beta(i+1) =&~ \frac{\sum_{t=1}^{T}\sum_{n=1}^{N} \left\{ \varpi_{n,t}(i)\big[|\gamma_{n,t}(i)|^2+\tau^{\gamma}_{n,t}(i)\big]\right\}}{\sum_{t=1}^{T}\sum_{n=1}^{N} \varpi_{n,t}(i)}, \label{equ:EM_beta} \\
    p_a(i+1) =&~ \frac{1}{N} \sum_{n=1}^{N}\frac{\overleftarrow{q}_{n,1}(i)\overrightarrow{q}_{n,1}(i)\overleftarrow{p}_{n,1}(i)}{(1-\overleftarrow{q}_{n,1}(i)) (1-\overrightarrow{q}_{n,1}(i))(1-\overleftarrow{p}_{n,1}(i))+\overleftarrow{q}_{n,1}(i)\overrightarrow{q}_{n,1}(i)\overleftarrow{p}_{n,1}(i)}, \label{equ:EM_pa} \\
    p_{10}(i+1) =&~ \frac{\sum_{t=2}^{T}\sum_{n=1}^{N}\Big(\mathbb{E}[\lambda_{n,t-1}]-\mathbb{E}[\lambda_{n,t-1}\lambda_{n,t}]\Big)}{\sum_{t=2}^{T}\sum_{n=1}^{N}\mathbb{E}[\lambda_{n,t-1}]}. \label{equ:EM_p10}
\end{align}
%\textcolor[rgb]{0.00,0.07,1.00}{
After obtaining the updated probabilities $p_a(i+1)$ and $p_{10}(i+1)$, we can simply obtain the updated probabilities $p_{01}(i+1) = p_a(i) p_{10}(i)/(1-p_a(i))$, $p_{11}(i+1) = 1-p_{10}(i+1), p_{00}(i+1) = 1-p_{01}(i+1)$ and finally complete the update of the transition probability matrix in the Markov chain.
To make the work self-contained, the detailed derivations of the above equations (\ref{equ:EM_tau_w}) to (\ref{equ:EM_p10}) are all provided in the Appendix. In addition, the explicit expressions of $\mathbb{E}[\lambda_{n,t-1}]$ and $\mathbb{E}[\lambda_{n,t-1}\lambda_{n,t}]$ are given as (\ref{equ:E_lamb_t-1}) and (\ref{equ:E_lamb_tt-1}) in the Appendix. To enable the EM algorithm to converge to the global maximum, we use a judicious initialization of the hyperparameters \cite{Moon_1996_SPM}:
\begin{align}
    \sigma^2_w(1) &= \frac{||\mathbf{Y}||^2_F}{(\text{SNR}^0+1)LT}, \label{equ:EM_ini_sigma} \\
    p_a(1) &= \frac{L}{N} \left\{ \max_{c>0} \frac{1-\frac{2N}{L}[(1+c^2\Phi(-c)-c\phi(c))]}{1+c^2-2[(1+c^2)\Phi(-c)-c\phi(c)]} \right\}, \label{equ:EM_ini_pa} \\
    \beta(1) &= \frac{||\mathbf{Y}||_F^2-LT\sigma_w(1)}{||A||_F^2p_aT}, \label{equ:EM_ini_beta} \\
    p_{10}(1) &= p_a(1), \label{equ:EM_ini_p10}
\end{align}
where $\Phi(\cdot)$ and $\phi(\cdot)$ denote the cumulative distribution function and the probability distribution function of the standard normal distribution, respectively. Since the knowledge of $\text{SNR}$ may not be available, the appropriate initial value of $\text{SNR}^0$ is in need. %\textcolor[rgb]{0.00,0.07,1.00}{
The conventional works usually adopt the value $\text{SNR}^0 = 20\text{dB}$ \cite{Vila_2013_TSP}, but we find that this setting can usually lead to a bad local optimal point for the EM-HyGAMP-DCS algorithm under our system setting. Though the suitable value of $\text{SNR}^0$ can be found by exhaustive search method, this approach needs to collect lots of data with known user activities and user channels, which is inefficient. To tackle this problem, we will first introduce the analysis tool of SE in the next section and then show that the SE can be an effective tool to find the suitable value $\text{SNR}^0$ for the specific system.
Finally, we outline the whole procedure of the EM-HyGAMP-DCS algorithm in Algorithm \ref{Alg:EM-SP-HyGAMP-DCS}.

\begin{algorithm}[t] \footnotesize
\caption{EM-HyGAMP-DCS for joint user activity detection and channel estimation}\label{Alg:EM-SP-HyGAMP-DCS}
\begin{algorithmic}[1]
\REQUIRE Pilot matrix $\mathbf{A}$, received signals $\mathbf{Y}$, tolerance $\varepsilon$, and maximum iteration $I_{max}$
\ENSURE MMSE estimate $\widehat{\mathbf{X}}$
\STATE \textbf{Initialize}
\STATE $i \leftarrow 1$
\STATE $\sigma^2_w(1)$, $p_a(1)$, $\beta(1)$, and $p_{10}(1)$ are initialized as (\ref{equ:EM_ini_sigma})--(\ref{equ:EM_ini_p10}).
\STATE $\forall n,t:~ \widehat{x}_{n,t}(i) = 0$, $\tau^x_{n,t}(i) = \mathbb{V}[x_{n,t}] = p_a(1) \beta(1)$, $\overrightarrow{p}_{n,t}(i) = p_a(i)$.
\STATE $\forall l,t:~ \widehat{s}_{l,t}(i-1) = 0$.
\STATE $\forall n:~ \overrightarrow{q}_{n,1}(1) = p_a(i)$.
\REPEAT
    \STATE \{HyGAMP-DCS estimation\}
    \STATE Perform line 8-24 in Algorithm \ref{Alg:SP-HyGAMP-DCS} with updated $\sigma_w(i)$, $p_a(i)$, $\beta(i)$, and $p_{10}(i)$.
    \STATE \{Hyperparameter update\}
    \STATE $\sigma^2_w(i+1)$, $\beta(i+1)$, $p_a(i+1)$ and $p_{10}(i+1)$ are updated as (\ref{equ:EM_tau_w})--(\ref{equ:EM_p10}).
    \STATE $i \leftarrow i+1$
\UNTIL{$i > I_{max}$ or $\frac{||\widehat{\mathbf{X}}(i+1)-\widehat{\mathbf{X}}(i)||_F^2}{||\widehat{\mathbf{X}}(i)||_F^2} \le \varepsilon$}.
\end{algorithmic}
\end{algorithm}

%\begin{align}\label{equ:EM_pa}
%    p_a(i+1) =&~ \frac{1}{N} \sum_{n=1}^{N}\frac{\overleftarrow{q}_{n,1}(i)\overrightarrow{q}_{n,1}(i)\overleftarrow{p}_{n,1}(i)}{(1-\overleftarrow{q}_{n,1}(i)) (1-\overrightarrow{q}_{n,1}(i))(1-\overleftarrow{p}_{n,1}(i))+\overleftarrow{q}_{n,1}(i)\overrightarrow{q}_{n,1}(i)\overleftarrow{p}_{n,1}(i)},
%\end{align}
%
%\begin{align}\label{equ:EM_p10}
%    p_{10}(i+1) &= \frac{\sum_{t=1}^{T-1}\sum_{n=1}^{N}\Big(\mathbb{E}[\lambda_{n,t}]-\mathbb{E}[\lambda_{n,t}\lambda_{n,t+1}]\Big)}{\sum_{t=1}^{T-1}\sum_{n=1}^{N}\mathbb{E}[\lambda_{n,t}]}
%\end{align}
%
%\begin{align}\label{equ:EM_beta}
%    \beta(i+1) &= \frac{\sum_{t=1}^{T}\sum_{n=1}^{N} \varpi_{n,t}(i)\Big[|\gamma_{n,t}(i)|^2+\tau^{\gamma}_{n,t}(i)\Big]}{\sum_{t=1}^{T}\sum_{n=1}^{N} \varpi_{n,t}(i)}
%\end{align}

\section{Performance and Complexity Analysis}\label{sec:performance_analysis}

In this section, we first analyze the performance of the proposed algorithms by the SE in the asymptotic regime, where $L, N \to \infty$ but $\frac{L}{N}$ is fixed. It is noted that the SE can not only be used to decide the setting on the pilot length $L$ and the number of joint detection frames $T$, but also play an important role of finding the appropriate hyperparameter initialization for the EM-HyGAMP-DCS algorithm. Then we illustrate its computational efficiency.

\subsection{Performance Analysis Using State Evolution}\label{sec:SE_equation}

SE is a common framework to analyze the estimation performance of the AMP-based algorithms in the asymptotic regime. When the condition that the pilot matrix has i.i.d. sub-Gaussian elements is satisfied, the SE can accurately track the MSE of estimator $\widehat{\mathbf{X}}(i)$ in large-scale problems. To simplify the SE analysis, we consider that HyGAMP-DCS has scalar variance in each frame $t$. These scalar variances of the algorithm \ref{Alg:SP-HyGAMP-DCS} can be rewritten as
\begin{align}
    {\tau}^{x}_{t}(i) \approx \bar{\tau}^{x}_{t}(i) &= \frac{1}{N}\sum_{n=1}^{N} \tau^{x}_{n,t}(i), \forall t, \label{equ:V_x_scalar} \\
    {\tau}^{p}_{t}(i) \approx \bar{\tau}^{p}_{t}(i) &= \frac{N}{L}{\tau}^{x}_{t}(i), \forall t, \label{equ:V_p_scalar} \\
    {\tau}^{z}_{t}(i) \approx \bar{\tau}^{z}_{t}(i) &= \frac{{\tau}^{p}_t(i)\sigma^{2}_{w}}{{\tau}^{p}_t(i)+\sigma^{2}_{w}}, \forall t, \label{equ:V_z_scalar} \\
    {\tau}^{s}_{t}(i) \approx \bar{\tau}^{s}_{t}(i) &= \frac{1}{{\tau}^{p}_{t}(i)}\left(1-\frac{{\tau}^{z}_{t}(i)}{{\tau}^{p}_{t}(i)}\right), \forall t, \label{equ:V_s_scalar} \\
    {\tau}^{r}_{t}(i) \approx \bar{\tau}^{r}_{t}(i) &= \frac{1}{{\tau}^{s}_{t}(i)}, \forall t, \label{equ:V_r_scalar}
\end{align}
then $\tau^{x}_{n,t}(i+1)$ is also calculated by (\ref{equ:V_x}).

Following the assumptions of the SE for the AMP-based algorithms in \cite{Rangan_2011_ISIT}, the asymptotic MSE of the estimator $\widehat{x}_{n,t}(i)$ is identical to $\mathbb{E}[\tau^{x}_{t}(i)]$ whose expectation is performed on $\widehat{r}_{n,t}(i)$.
Specifically, for the random variable $x_0 \sim p_X(x_0)$ with $p_X(x_0) = (1-\rho_0)\delta(x_0) + \rho_0\mathcal{CN}(x_0;0,\beta)$, the variable $\widehat{r}_{0}(i)$ in each iteration $i$ can be modeled as
\begin{align}
    \widehat{r}_{0}(i) &= x_{0} + n^{r}_{0}(i),
\end{align}
where $n^{r}_{0}(i) \sim \mathcal{CN}(0,\tau^{r}_{0}(i))$ is the corrupting complex Gaussian noise being independent to $x_{0}$. Under the proposed algorithm with scalar variance, we have
\begin{align}\label{equ:SE_rx}
    \tau^{r}_{0}(i) &= \frac{(\tau^{p}_{0}(i))^2}{\tau^{p}_{0}(i)-\tau^{z}_{0}(i)} = \sigma^2_{w} + \frac{N}{L}\mathbb{E}[\tau^{x}_{0}(i)], %\notag \\
    %%\quad &= \sigma^2_{w} + \tau^{p}_{0}(i) \notag \\
    %\quad &= \sigma^2_{w} + \frac{N}{L}\mathbb{E}[\tau^{x}_{0}(i)],
\end{align}
where $\tau^{x}_{0}(i)$ is calculated by (\ref{equ:V_x}) and the expectation operates on both $x_{0}$ and $n^{r}_{0}(i-1)$. The explicit expression of the expectation is given as
\begin{align}\label{equ:E_V_x}
    \mathbb{E}[\tau^{x}_{0}(i)] &= \iint \tau^{x}_{0}(i) p_X(x_0) p(n^r_0(i-1)) ~dx_0dn^r_0 \notag \\
    \quad &= \frac{\rho_0 \beta \tau^{r}_{0}(i-1)}{\beta+\tau^r_{0}(i-1)} + \frac{\rho_0 \beta^2 \left(1-\psi(\beta/\tau^{r}_{0}(i-1))\right)}{\beta+\tau^{r}_{0}(i-1)}
\end{align}
where
\begin{align}\label{equ:psi}
    \psi(b) &= \int_{0}^{\infty} \frac{s\exp(-s)}{1+(\rho^{-1}_0-1)(1+b)\exp(-bs)} ~ds.
\end{align}
Then the asymptotic MSE of $\widehat{\mathbf{X}}(i)$ in each iteration $i$ can be given as
\begin{align}\label{equ:V_E_x_t}
    \mathbb{E}[\tau^{x}(i)] &= \mathbb{E}\Bigg[\frac{1}{NT}\sum_{n=1}^{N}\sum_{t=1}^{T}\tau^{x}_{n,t}(i)\Bigg] = \frac{1}{NT}\sum_{n=1}^{N}\sum_{t=1}^{T}\mathbb{E}[\tau^{x}_{n,t}(i)]. %\notag \\
    %\quad &= \frac{1}{NT}\sum_{n=1}^{N}\sum_{t=1}^{T}\mathbb{E}[\tau^{x}_{n,t}(i)].
\end{align}

However, the active probability $\rho_{0}$ is updated at the MP part in each iteration and it is hard to give the explicit expression of the marginal probability on $\rho_0$. Thus, we resort to the Monte Carlo simulations to obtain $\rho_{n,t} = \overrightarrow{p}_{n,t}(i)$ for each user $n$ in frame $t$ at each iteration and then substitute it into (\ref{equ:V_E_x_t}). %\textcolor[rgb]{0.00,0.07,1.00}{
Similarly, the performance of the EM-HyGAMP-DCS algorithm can also be predicted by following the above procedure where the statistical parameter set $\pmb{\vartheta}$ in each iteration are replaced by the learned one $\pmb{\vartheta}(i)$.%}
%In specific, the hyperparameters in $\pmb{\vartheta}$ are also updated by simulations and then substituted into (\ref{equ:SE_rx}) and (\ref{equ:E_V_x}) to obtain the SE.

\begin{figure}[t]
  \centering
  \subfigure[EM-HyGAMP-DCS]
  {\includegraphics[width=.4\textwidth]{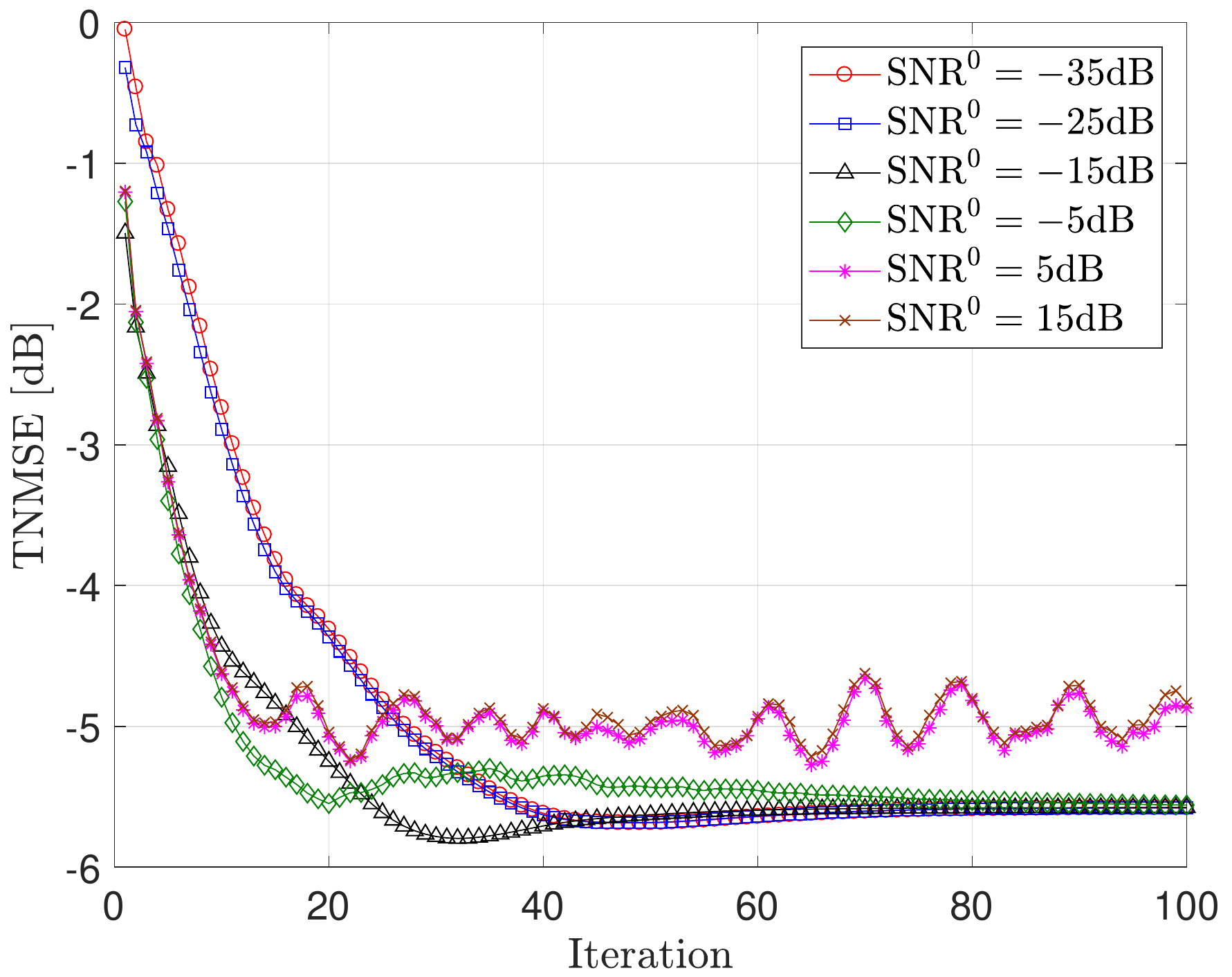}}
  \subfigure[The SE of EM-HyGAMP-DCS]
  {\includegraphics[width=.4\textwidth]{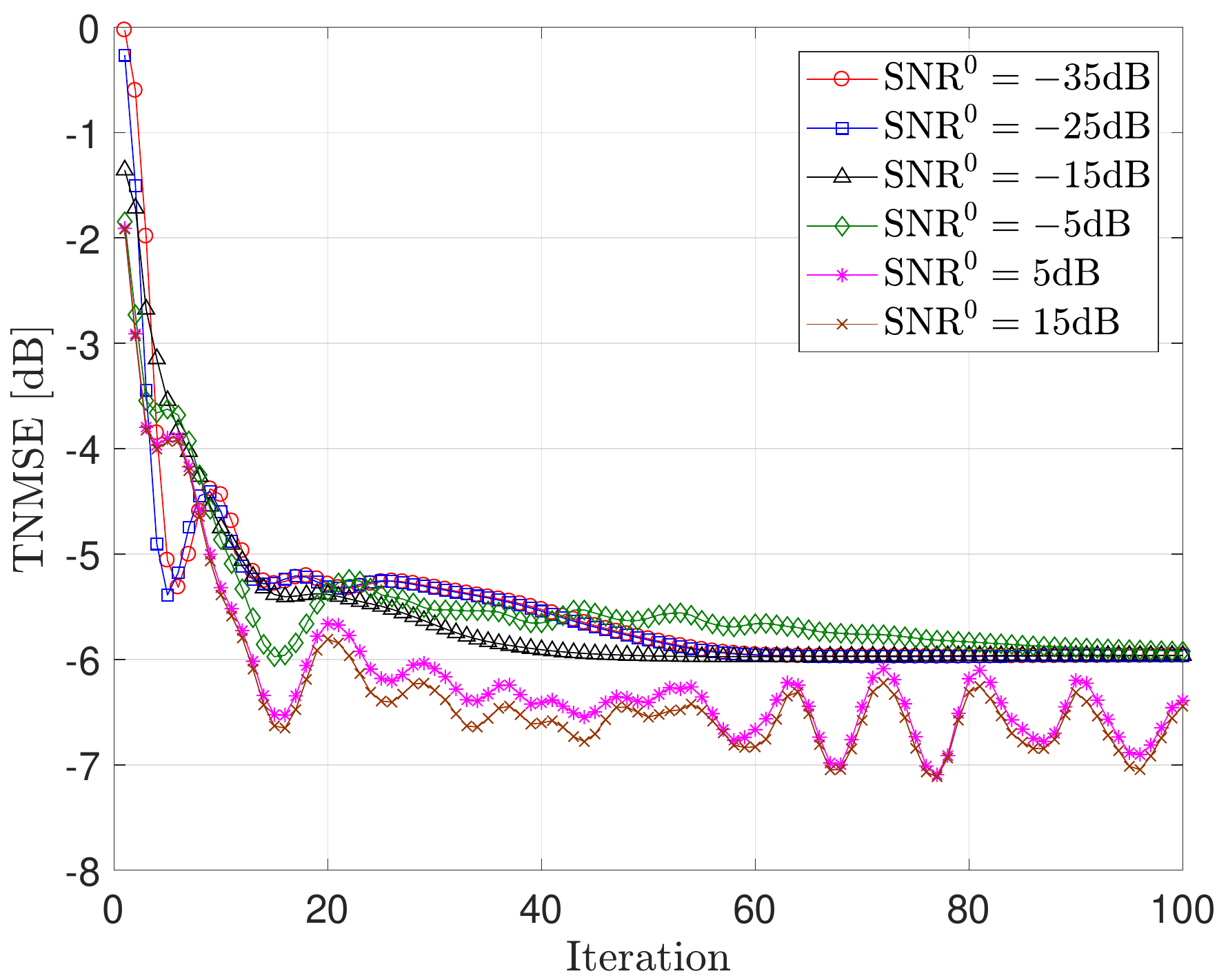}}
  \vspace{-0.3cm}
  %\textcolor[rgb]{0.00,0.07,1.00}{
  \caption{The impact of the value of $\text{SNR}^0$ on the recovery performance of EM-HyGAMP-DCS and the SE under the system setting $\text{SNR} = \frac{\beta}{\sigma_w^2} = -10 \text{dB}$ (The value $\text{SNR}^0 = -15 \text{dB}$ is appropriate).}\label{Fig:SE-Simu-SNR-m10dB}%}
  \vspace{-0.6cm}
\end{figure}

\begin{figure}[t]
  \centering
  \subfigure[EM-HyGAMP-DCS]
  {\includegraphics[width=.4\textwidth]{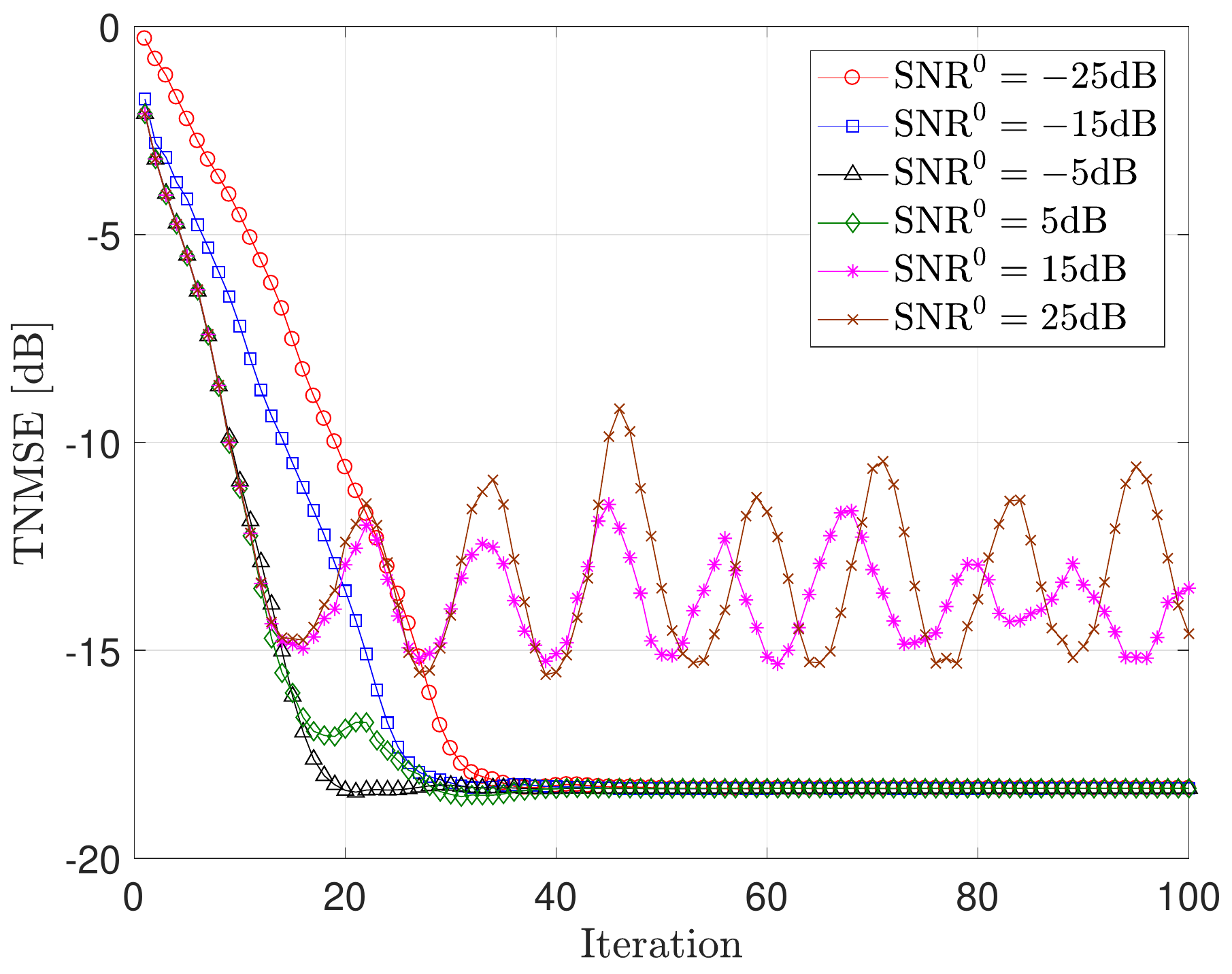}}
  \subfigure[The SE of EM-HyGAMP-DCS]
  {\includegraphics[width=.4\textwidth]{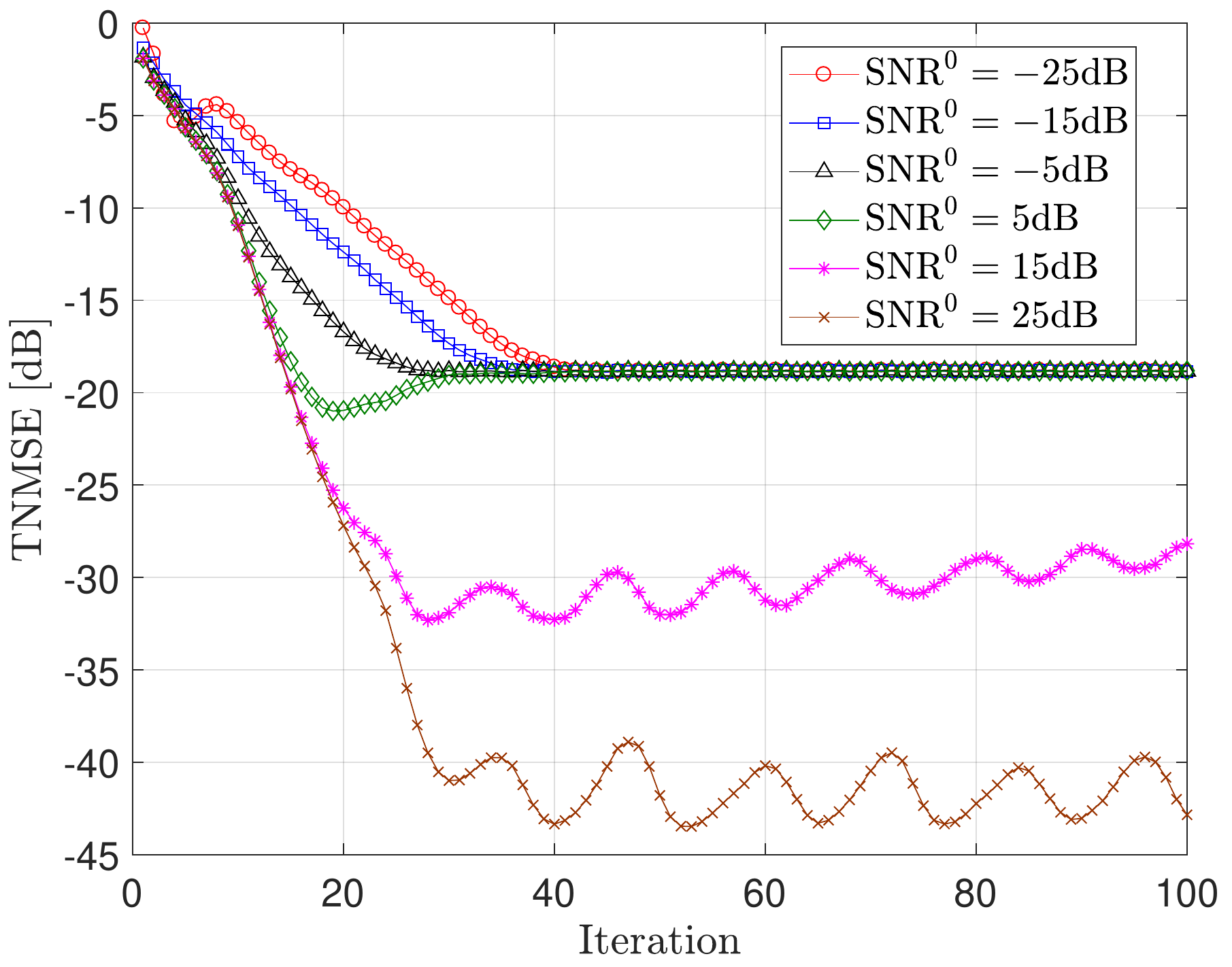}}
  \vspace{-0.3cm}
  %\textcolor[rgb]{0.00,0.07,1.00}{
  \caption{The impact of the value of $\text{SNR}^0$ on the recovery performance of EM-HyGAMP-DCS and the SE under the system setting $\text{SNR} = \frac{\beta}{\sigma_w^2} = 0 \text{dB}$ (The value $\text{SNR}^0 = -5 \text{dB}$ is appropriate).}\label{Fig:SE-Simu-SNR-0dB}
  \vspace{-0.8cm}
\end{figure}

%\textcolor[rgb]{0.00,0.07,1.00}{
From the simulation trials, we can find that the EM-HyGAMP-DCS algorithm easily converges to a bad local optimal point with an inappropriate selection on $\text{SNR}^0$. As usual, we can easily collect lots of signals $\{\mathbf{y}^t\}$ received at the BS, but the true channels $\{\mathbf{x}^0_{t}\}$ are hard to be obtained in advance, which hinders us to find the appropriate initialization based on empirical results. Fortunately, the SE can be utilized to find an appropriate value of $\text{SNR}^0$ for the initialization of the EM-HyGAMP-DCS algorithm in different systems. Based on the SE, we can try different values of $\text{SNR}^0$ and then select the suitable one that assures the EM-HyGAMP-DCS algorithm to converge to a satisfactory point. Fig. \ref{Fig:SE-Simu-SNR-m10dB} and Fig. \ref{Fig:SE-Simu-SNR-0dB} show the impact of the value of $\text{SNR}^0$ on the algorithm performance and the SE under the systems with $\text{SNR} = \frac{\beta}{\sigma_w^2} = -10 \text{~dB}$ and $\text{SNR} = \frac{\beta}{\sigma_w^2} = 0 \text{~dB}$, respectively.
Here, we use the time-averaged normalized MSE (TNMSE) as the performance metric, which is computed as
\begin{equation}\label{equ:TNMSE}
    \text{TNMSE} = 10 \log_{10} \left(\frac{||\widehat{\mathbf{X}}-\mathbf{X}||^2_F}{||\mathbf{X}||_F^2} \right).
\end{equation}

The numerical results in Fig. \ref{Fig:SE-Simu-SNR-m10dB} and Fig. \ref{Fig:SE-Simu-SNR-0dB} show that the performance of EM-HyGAMP-DCS is sensitive to the value of $\text{SNR}^0$.
We can observe that using $\text{SNR}^0 < \text{SNR}$ can always ensure EM-HyGAMP-DCS algorithm to converge to a satisfactory point but may have a slow convergence rate. For $\text{SNR}^0 \gg \text{SNR}$, the EM-HyGAMP-DCS algorithm usually cannot converge to a stationary point and thus achieve bad performance. Similarly, the SE by using $\text{SNR}^0 < \text{SNR}$ can converge to a stationary point. However, if $\text{SNR}^0 \ll \text{SNR}$, the SE has TNMSE fluctuation before it converges in the case of $\text{SNR} = -10\text{~dB}$. The SE also have a slower convergence rate when we set $\text{SNR}^0 \ll \text{SNR}$. When we have $\text{SNR}^0 \gg \text{SNR}$, the SE cannot converge to a stationary point though it seems to give a smaller TNMSE. For the $\text{SNR}^0$ being close to the true $\text{SNR}$, the SE is shown to be smooth without TNME fluctuation and has fast convergence rate. In summary, the SE can be an essential tool to select an appropriate value of $\text{SNR}^0$, which makes EM-HyGAMP-DCS achieve satisfactory performance and fast convergence.%}

%\textcolor[rgb]{0.00,0.07,1.00}{
Besides using the SE to find a suitable value of $\text{SNR}^0$, we can also find the appropriate initialization of the other hyperparameters by the SE. Thus, the SE is not only used to analyze the asymptotic performance of the AMP-based algorithms under different system settings, but also a powerful tool to find the appropriate hyperparameter initialization for the AMP-based algorithms incorporated with the EM techniques.%}

\subsection{Computational Complexity Analysis}

We compare the computational complexity of the proposed algorithms with those of the benchmark algorithms in Table \ref{table:CC}.
We consider five benchmarks which are the DCS-OMP algorithm \cite{Wang_2016_CL}, the PIA-ASP algorithm \cite{Du_2017_JSAC}, the GAMP algorithm \cite{Rangan_2011_ISIT}, the S-AMP algorithm \cite{Jiang_2021_TWC} and DCS-AMP algorithm \cite{Ziniel_2013_TSP}.
Note that DCS-OMP, PIA-ASP, GAMP, and S-AMP are frame-by-frame detection schemes, while DCS-AMP and the proposed HyGAMP-DCS and EM-HyGAMP-DCS are block-by-block detection schemes. Thus, for fair comparison, the number of the complex value multiplications at the each iteration $i$ normalized by the number of joint detection frames, $T$, denoted as $C_i$, is given.
Then the average computational complexity per frame of each algorithm can be obtained by $C = \sum_{i=1}^{I} C_i$ with $I$ being the iteration number.
%Here, the number of the complex multiplications at the $i$th iteration by normalizing the number of joint detection frames\footnote{\textcolor[rgb]{0.00,0.07,1.00}{DCS-OMP, PIA-ASP, GAMP, S-AMP all perform frame-by-frame detection, then the number of joint detection frames for them is $1$ and $C_i$s of these four algorithms are the computational complexities in the $i$th iteration for the estimation of each frame. For DCS-AMP, HyGAMP-DCS and EM-HyGAMP-DCS with block-by-block detection, we consider that each block contains $T$ consecutive frames and then the number of joint detection frames for them is $T$.}}, denoted as $C_i$, is used as the performance metric.
%For fair comparison, we evaluate the total computational complexity on joint activity detection and channel estimation for $T$ consecutive frames of each algorithm, which is obtained as $C = T\sum_{i=1}^{I}C_i$ with $I$ being the iteration number.}%\footnote{\textcolor[rgb]{0.00,0.07,1.00}{Specially, for S-AMP, we have $C = T\sum_{i=1}^{I}C_i + 8N(T-1)$.}}
%Note that DCS-OMP, PIA-ASP, GAMP and S-AMP all contains two levels of iterations as the outer iteration and the inner iteration.  These three algorithms and the S-AMP algorithm all perform joint user activity detection and channel estimation of each frame sequentially in the outer iteration, while DCS-AMP, HyGAMP-DCS and EM-HyGAMP-DCS perform user activity detection and channel estimation of all detected frames simultaneously.
We can find that the average computational complexities per frame of DCS-OMP and PIA-ASP are in the same order of $\mathcal{O}(I(p_a N)^3)$, since these two algorithms adopt least square (LS) estimation which performs matrix inversion operation in each iteration. By contrast, the average computational complexities per frame of the AMP-based algorithms are in the order of $\mathcal{O}(ILN)$.
Compared with the standard GAMP algorithm, the HyGAMP-DCS algorithm and the EM-HyGAMP-DCS algorithm have slightly higher computational complexity due to the additional message updates between the adjacent frames.
%Since $C_i$s of both HyGAMP-DCS and EM-HyGAMP-DCS are irrelevant to $T$, increasing the number of joint detection frames will not introduce additional computation costs.
Additionally, we can find that HyGAMP-DCS and EM-HyGAMP-DCS may have faster convergence than the GAMP algorithm from numerical results, meaning that the computational cost of these two proposed algorithms can be further reduced.

\begin{table*}[t]
  \centering
  \caption{Computational Complexity Comparison}\label{table:CC}
    \begin{tabular}{c|c|c}
    \hline
    \hline
    Algorithm & \makecell{Number of complex value multiplications at each iteration $i$ \\ normalized by the number of joint detection frames ($C_i$)}  & \makecell{The order of the average computational \\ complexity per frame $\mathcal{O}(C)$} \\
    \hline
    DCS-OMP & $LN + 2N + Li + 2Li^2 + i^3$ & $\mathcal{O}(I(p_aN)^3)$ \\
    PIA-ASP & $LN + 6N - 2s_p + L(s_p+j) + 6L(s_p+j)^2 + 3(s_p+j)^3$ & $\mathcal{O}(I(p_aN)^3)$ \\
    GAMP & $4LN + 9L + 16N$ & $\mathcal{O}(ILN)$ \\
    S-AMP & $4LN + 9L + 24N$ & $\mathcal{O}(ILN)$ \\
    DCS-AMP & $4I_0LN + 9L + 32N$ & $\mathcal{O}(II_0LN)$ \\
    HyGAMP-DCS & $4LN + 9L + 32N$ & $\mathcal{O}(ILN)$ \\
    EM-HyGAMP-DCS & $4LN + 9L + 45N$ & $\mathcal{O}(ILN)$ \\
    \hline
    \hline
	\end{tabular} \\
\vspace{0.2cm}
Note: $i$ indicates the iteration index, $j$ and $s_p$ indicate the index of sparsity level and quality information of the PIA-ASP algorithm, $I_0$ indicates the iteration number of the inner AMP estimation in the DCS-AMP algorithm.
\vspace{-0.6cm}
\end{table*}

\begin{remark}
The proposed HyGAMP-DCS algorithm has similar structure to the DCS-AMP algorithm proposed in \cite{Ziniel_2013_TSP}. %\textcolor[rgb]{0.00,0.07,1.00}{
However, the DCS-AMP algorithm in each outer iteration runs a complete AMP algorithm where the variables are firstly initialized and then iteratively updated for $I_0 \gg 1$ times \cite{Ziniel_2013_TSP}. Thanks to the $I_0$-iteration AMP module, DCS-AMP may needs only several outer iterations for messages exchange between the adjacent frames to update the activity likelihoods. In each iteration of HyGAMP-DCS, the variables in the GAMP part are only updated once based on the estimated results in the last iteration.
Note that the computational cost of DCS-AMP and HyGAMP-DCS mainly results from the calculation in the AMP estimation. Therefore, the whole computational complexity of DCS-AMP is usually higher than that of HyGAMP-DCS, though DCS-AMP only has several outer iterations.
\end{remark}

%\vspace{-2cm}
\section{Simulation Results}\label{sec:simulation}

In this section, we give the simulation results of our proposed algorithm in the temporally-correlated massive access system. To show the advantage of our proposed algorithm, the conventional DCS-based algorithm of PIA-ASP \cite{Du_2017_JSAC}, the standard GAMP algorithm \cite{Rangan_2011_ISIT} and the S-AMP algorithm\footnote{From our simulation trials, we find that the S-AMP algorithm has very similar performance to the SI-aided MMV-AMP algorithm \cite{Wang_2021_ISIT}. Thus, we only show the performance of the S-AMP algorithm in this section.} \cite{Jiang_2021_TWC} are also evaluated as the benchmarks.

We consider the simulation scenario where there are $N=10^3$ users.
%We consider that the power control strategy is employed at the device side as \cite{Senel_2018_TCOM} does, so that we can obtain $\beta_{1,1} = \beta_{1,2} = \dots = \beta_{N,T} = \beta$. Then
The number of consecutive frames for joint detection is set as $T=4$ and we set $p_{10} = 0.25$, if not specified.
The performance metric used for the user activity detection is defined as the time-averaged activity error ratio (TAER), which is calculated as
\begin{equation}\label{equ:TAER}
    \text{TAER} = \frac{\sum_{t=1}^{T}[N_{m}(t)+N_{f}(t)]}{NT},
\end{equation}
where $N_{m}(t)$ is the number of missed detected users in the $t$th frame and $N_{f}(t)$ is the number of false alarm users in the $t$th frame.
All the numerical results here are obtained by averaging over $10^4$ channel realizations.

%\begin{figure}[t]
%  \centering
%  \includegraphics[width=.5\textwidth]{J_Fig_Err_SNR.eps}
%  \caption{The performance of user activity detection in terms of the TAER versus SNR}\label{Fig:TAER}
%\end{figure}
%
%\begin{figure}[t]
%  \centering
%  \includegraphics[width=.5\textwidth]{Fig_TNMSE_ICC2.pdf}
%  \caption{The performance of channel estimation in terms of TNMSE versus SNR}\label{Fig:TNMSE}
%\end{figure}

\begin{figure}[t]
  \centering
  \begin{minipage}[t]{.45\textwidth}
    \center
    \includegraphics[width=\textwidth]{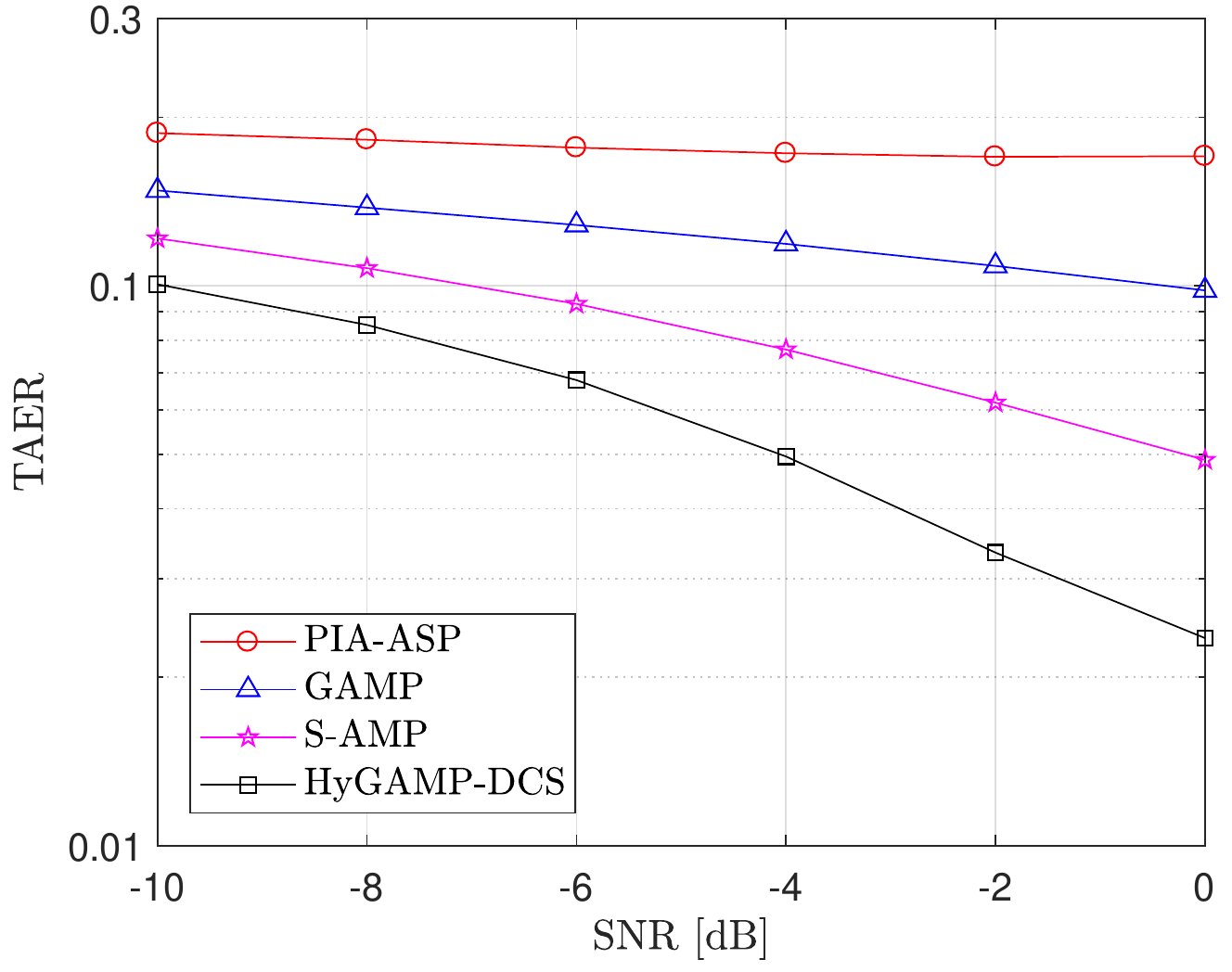}
    \vspace{-1.4cm}
    \caption{The impact of SNR on the user activity detection performance.}\label{Fig:TAER}
  \end{minipage}
  \begin{minipage}[t]{.44\textwidth}
    \center
    \includegraphics[width=\textwidth]{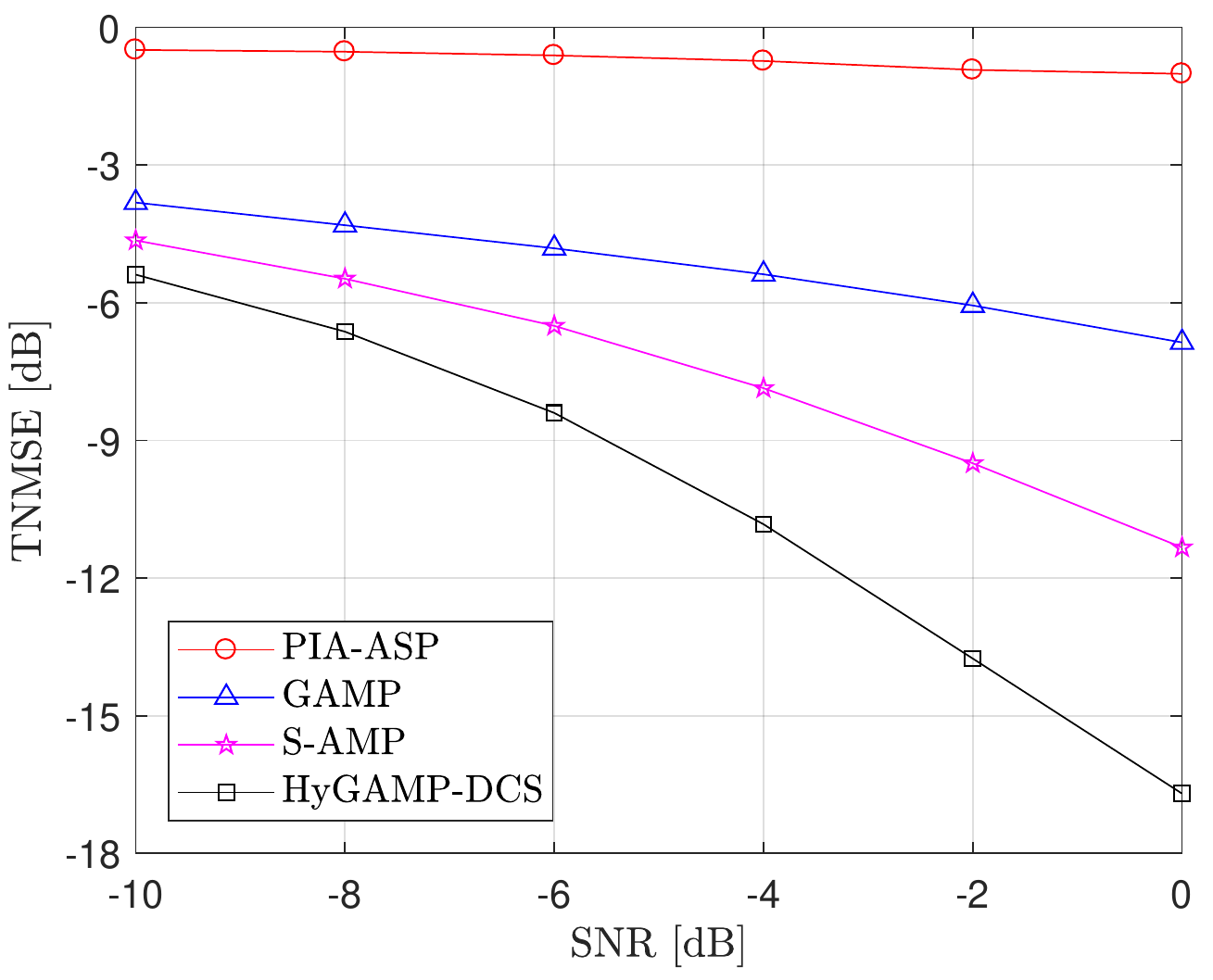}
    \vspace{-1.4cm}
    \caption{The impact of SNR on the channel estimation performance.}\label{Fig:TNMSE}
  \end{minipage}
\vspace{-0.7cm}
\end{figure}

\subsection{Comparison with the Benchmarks}

We evaluate the performance of the proposed HyGAMP-DCS algorithm for user activity detection and channel estimation in Fig. \ref{Fig:TAER} and Fig. \ref{Fig:TNMSE}. We set $L = 300$ and $p_a = 0.2$. One user is detected to be active when the power of its estimated channel is larger than the previously selected threshold $\varrho$, i.e., $|\widehat{x}_{n,t}|^2 > \varrho$, and the value of $\varrho$ is the same for these three schemes. The result shows that the three AMP-based algorithms can all outperform the PIA-ASP algorithm, meaning that the prior knowledge of channels can be exploited to greatly improve the performance. It is also observed that the gaps between the PIA-ASP algorithm and the three AMP-based algorithms become larger when SNR increases, which implies that the AMP-based algorithm can acquire larger performance gain than the greedy algorithm in high SNR scenario. By exploiting the temporal correlation, the S-AMP algorithm and the proposed HyGAMP-DCS algorithm can achieve better performance than the GAMP algorithm. Moreover, the HyGAMP-DCS algorithm has further improved performance over S-AMP by performing bidirectional message propagation at the cost of the increased detection delay.

%Then the performance of the channel estimation in terms of the TNMSE versus SNR is shown in Fig. \ref{Fig:TNMSE}. It is obvious that the statistical information of the channels can help reduce the estimation error. And the gaps between the PIA-ASP algorithm and the two GAMP-based algorithms become larger when SNR increases, which implies that the GAMP-based algorithm can acquire larger performance gain than the greedy algorithm in higher SNR. By considering the dynamic support structure, the proposed algorithm can refine the soft user activity information to improve the recovery performance of the GAMP estimation.

\begin{figure}[t]
  \centering
  \begin{minipage}[t]{.45\textwidth}
    \center
    \includegraphics[width=\textwidth]{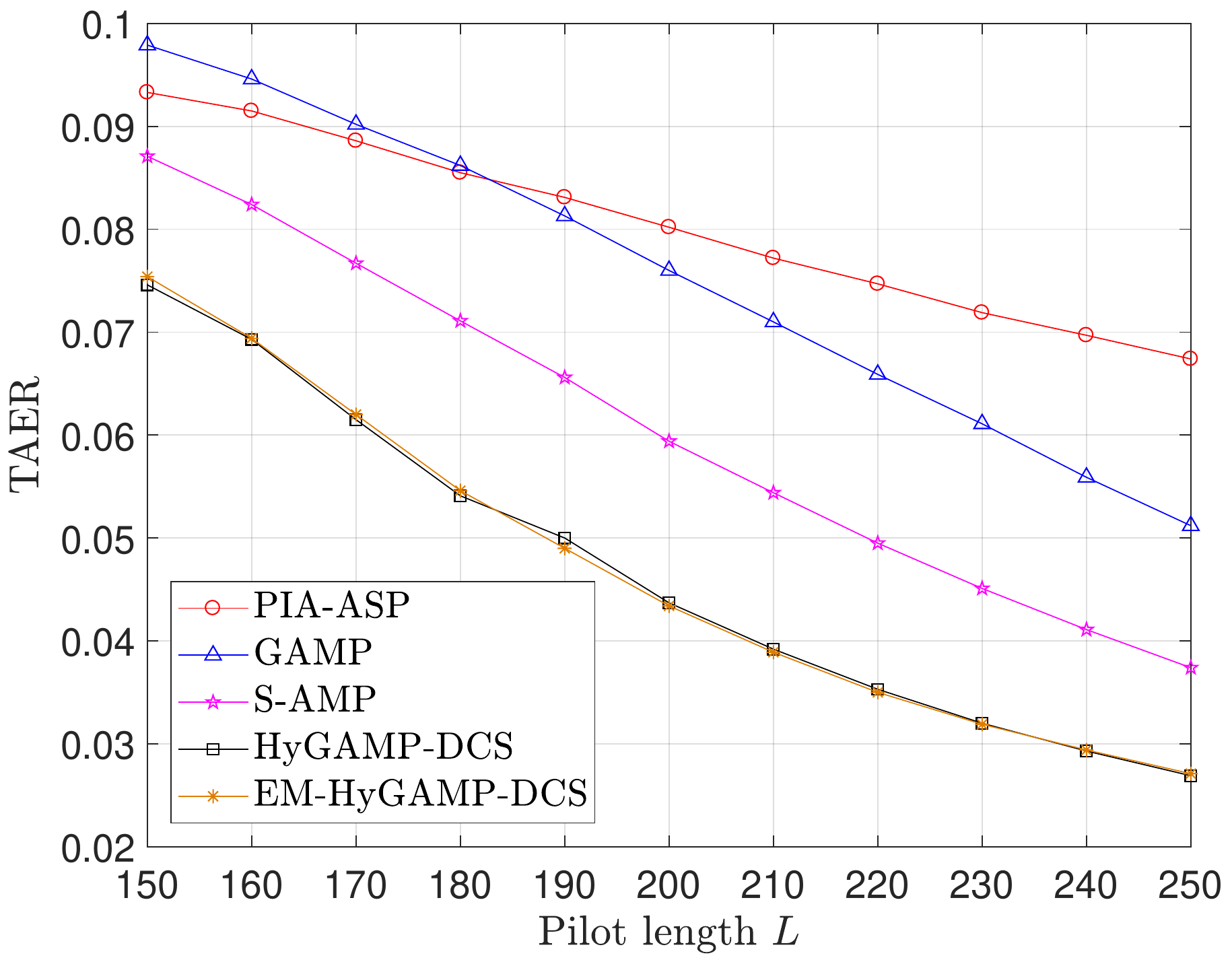}
    \vspace{-1.4cm}
    \caption{The impact of pilot length $L$ on the user activity detection performance.}\label{Fig:TAER_L}
  \end{minipage}
  \begin{minipage}[t]{.45\textwidth}
    \center
    \includegraphics[width=\textwidth]{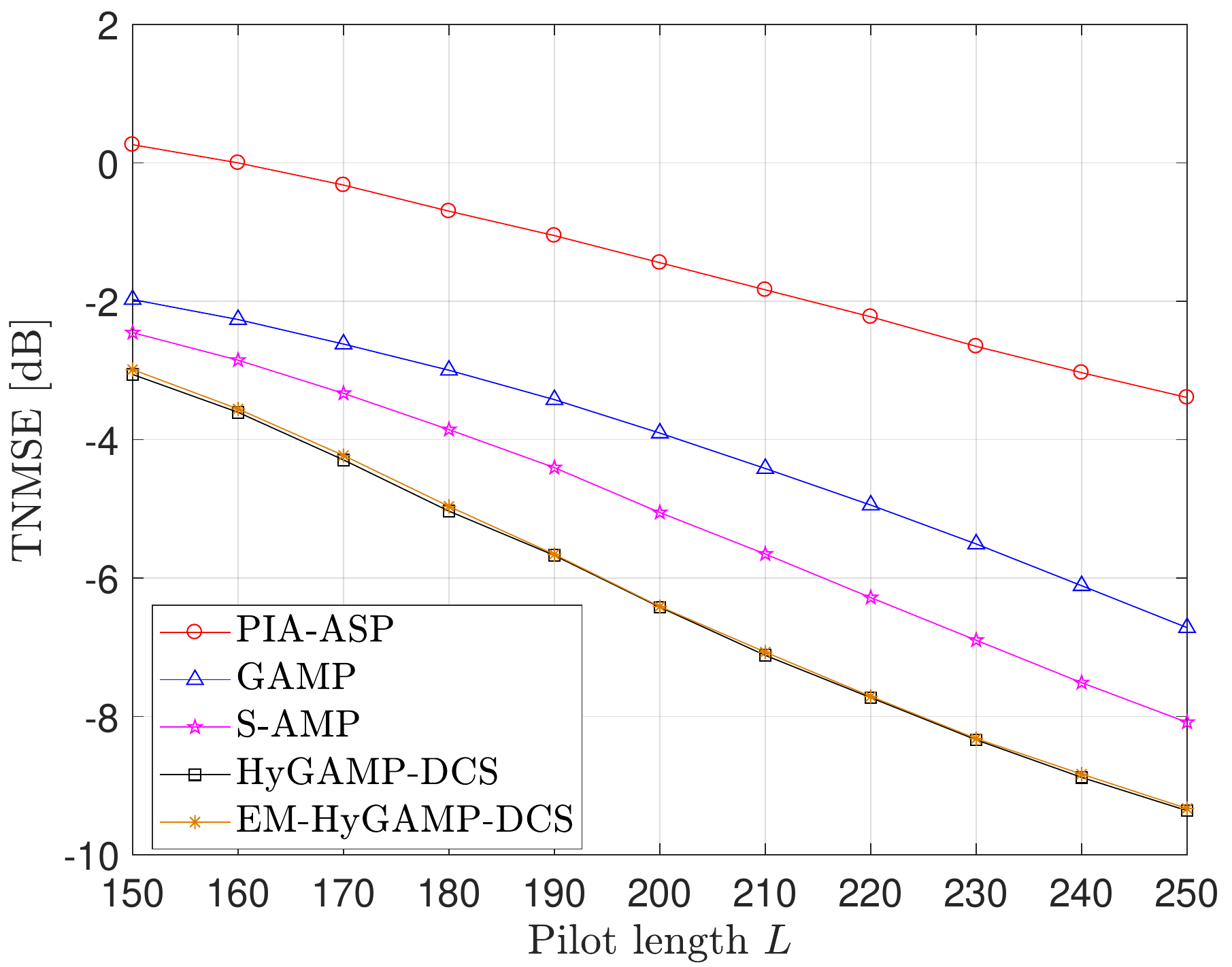}
    \vspace{-1.4cm}
    \caption{The impact of pilot length $L$ on the channel estimation performance.}\label{Fig:TNMSE_L}
  \end{minipage}
\vspace{-0.7cm}
\end{figure}

The impact of the pilot length $L$ on the performance of the proposed algorithms and the benchmarks is evaluated in Fig. \ref{Fig:TAER_L} and Fig. \ref{Fig:TNMSE_L}. We set $p_a = 0.1$ and $\text{SNR} = -10\text{dB}$. We observe that the AMP-based algorithms all have larger performance gaps to the PIA-ASP algorithm when $L$ increases, which indicates the essential role of the channel statistics for performance improvement. It is also observed that the PIA-ASP algorithm can outperform the GAMP algorithm in the user activity detection performance when $L$ is smaller than 190, though the GAMP algorithm always has better channel estimation performance. When $L$ keeps increasing, the GAMP algorithm can outperform the PIA-ASP algorithm, which means that exploiting the temporal correlation provides limited performance gain for the greedy-based algorithm. The EM-HyGAMP-DCS algorithm is shown to achieve similar performance to the HyGAMP-DCS algorithm even if the perfect system statistics are unknown, implying that the EM-HyGAMP-DCS algorithm can be suitable to the practical scenarios.

\begin{figure}[t]
 %   \vspace{-0.3cm}
  \centering
  \begin{minipage}[t]{.45\textwidth}
    \center
    \includegraphics[width=\textwidth]{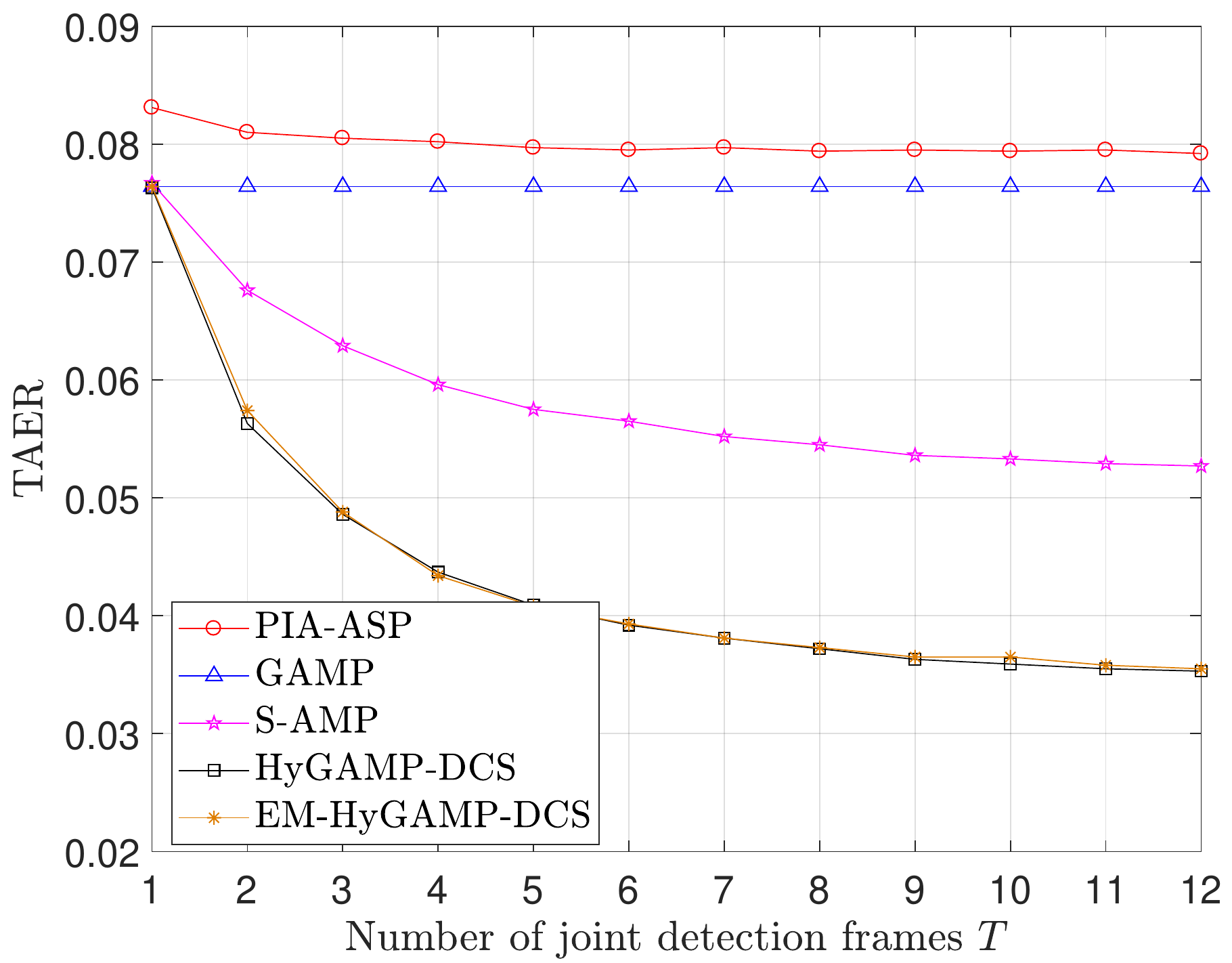}
    \vspace{-1.4cm}
    \caption{The impact of the number of joint detection frames, $T$, on the user activity detection performance.}\label{Fig:TAER_T}
  \end{minipage}
  \begin{minipage}[t]{.435\textwidth}
    \center
    \includegraphics[width=\textwidth]{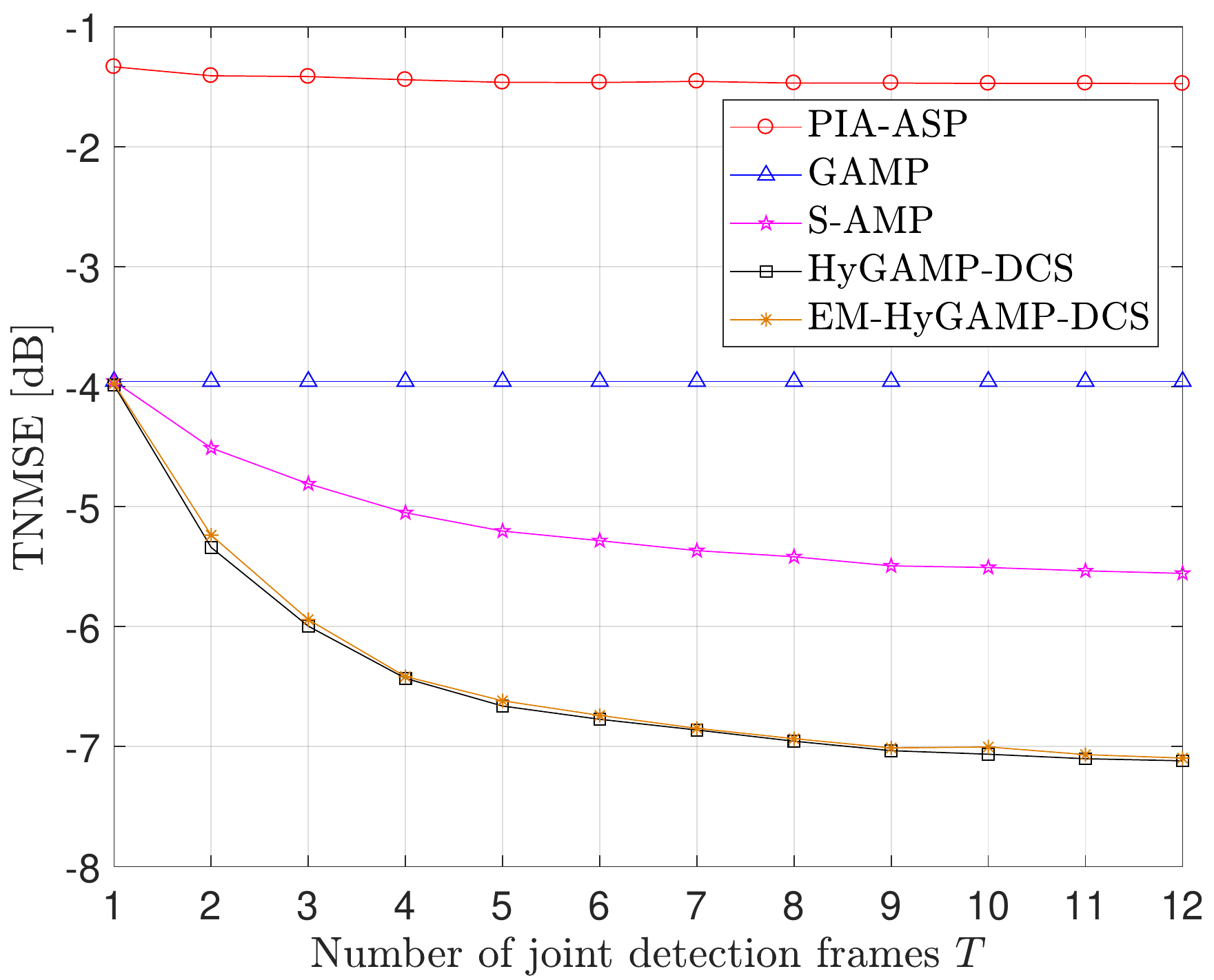}
    \vspace{-1.4cm}
    \caption{The impact of the number of joint detection frames, $T$, on the channel estimation performance.}\label{Fig:TNMSE_T}
  \end{minipage}
  \vspace{-0.4cm}
\end{figure}

%\begin{figure}[t]
%  \centering
%  \includegraphics[width=.5\textwidth]{Fig_TAER_T_ICC2.pdf}
%  \caption{TAER versus T}\label{Fig:TAER_T}
%\end{figure}
%
%\begin{figure}[t]
%  \centering
%  \includegraphics[width=.5\textwidth]{Fig_TNMSE_T_ICC2.pdf}
%  \caption{TNMSE versus T}\label{Fig:TNMSE_T}
%\end{figure}

Fig. \ref{Fig:TAER_T} and Fig. \ref{Fig:TNMSE_T} show the impact of the number of joint detection frames $T$, on the performance of the evaluated algorithms. We set $L=200$, $p_a=0.1$ and $\text{SNR}=-10\text{dB}$. Both the TAER and TNMSE of the DCS-based algorithms is reduced when we increases $T$. Moreover, the TAERs and TNMSEs of the S-AMP algorithm, the HyGAMP-DCS algorithm and the EM-HyGAMP-DCS algorithm have sharp reductions when $T$ increases and $T<4$.
%The reason is that the ratio of the frames where the estimations can be refined with the double-sided messages to the total number of frames increases with a larger $T$.
However, the performance improvement diminishes as $T$ keep increasing, and eventually the performance saturates. As such, we can set a moderate value $T$ to take a compromise between performance and detection delay.

\begin{figure}[t]
 %   \vspace{-0.3cm}
  \centering
  \begin{minipage}[t]{.45\textwidth}
    \center
    \includegraphics[width=\textwidth]{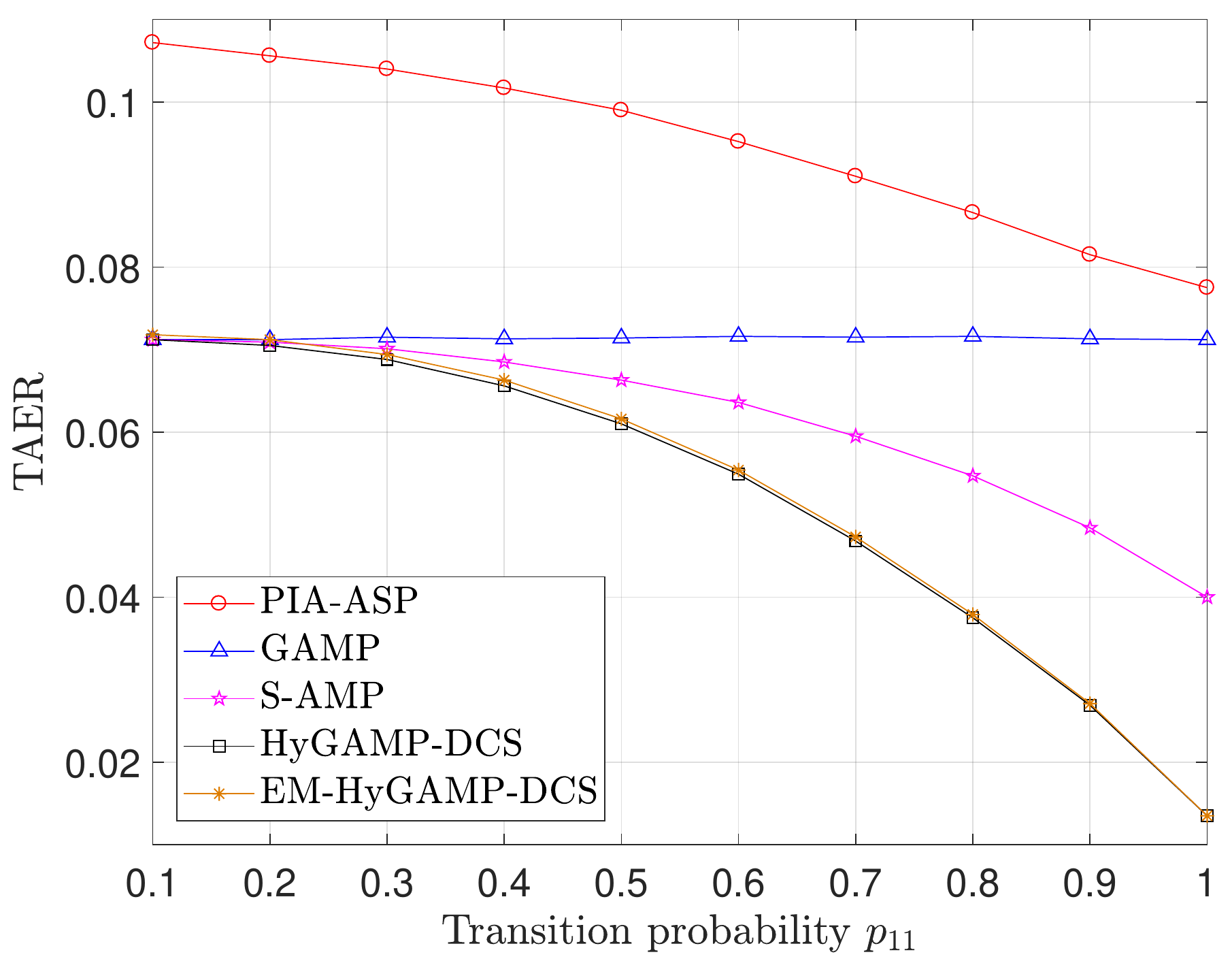}
    \vspace{-1.4cm}
    \caption{The impact of transition probability $p_{11}$ on the user activity detection performance.}\label{Fig:TAER_p11}
  \end{minipage}
  \begin{minipage}[t]{.45\textwidth}
    \center
    \includegraphics[width=\textwidth]{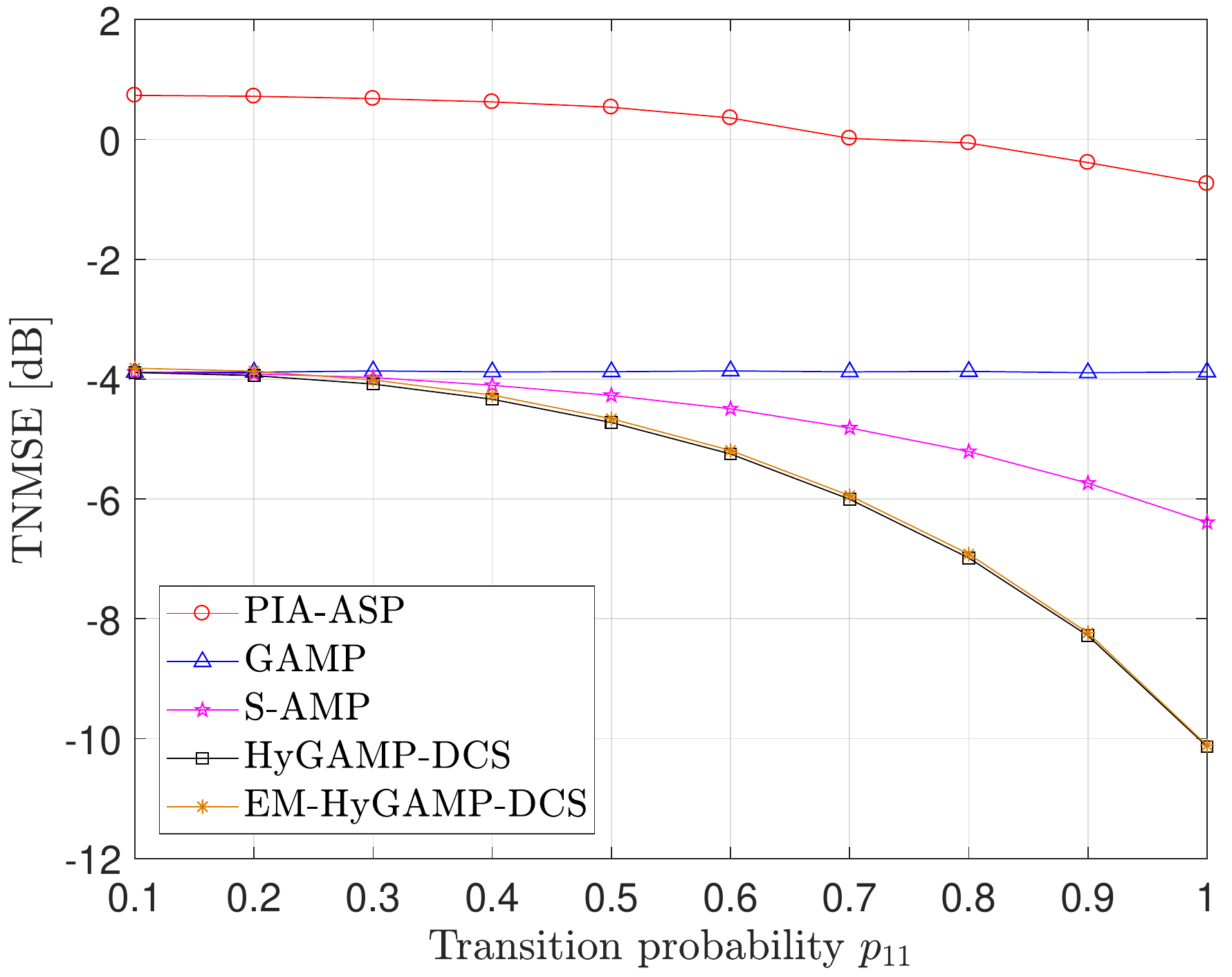}
    \vspace{-1.4cm}
    \caption{The impact of transition probability $p_{11}$ on the channel estimation performance.}\label{Fig:TNMSE_p11}
  \end{minipage}
  \vspace{-0.5cm}
\end{figure}

Fig. \ref{Fig:TAER_p11} and Fig. \ref{Fig:TNMSE_p11} depict the impact of the transition probability $p_{11}$ on the algorithm performance under fixed active probability $p_a = 0.1$. We also set $L=200$ and $\text{SNR} = -10\text{~dB}$. It is observed that DCS-based algorithms all have smaller TAER and TNMSE with stronger temporal correlation and their performance is greatly enhanced when $p_{11}$ approaches 1. These results demonstrate that exploiting the temporal correlation can greatly improve the user activity and channel estimation performance. Furthermore, our proposed algorithms can significantly outperform the benchmarks especially in the scenario with strong temporal correlation, which validates the superior performance of the proposed algorithms.

\begin{figure}[t]
 %   \vspace{-0.3cm}
  \centering
  \begin{minipage}[t]{.45\textwidth}
    \center
    \includegraphics[width=\textwidth]{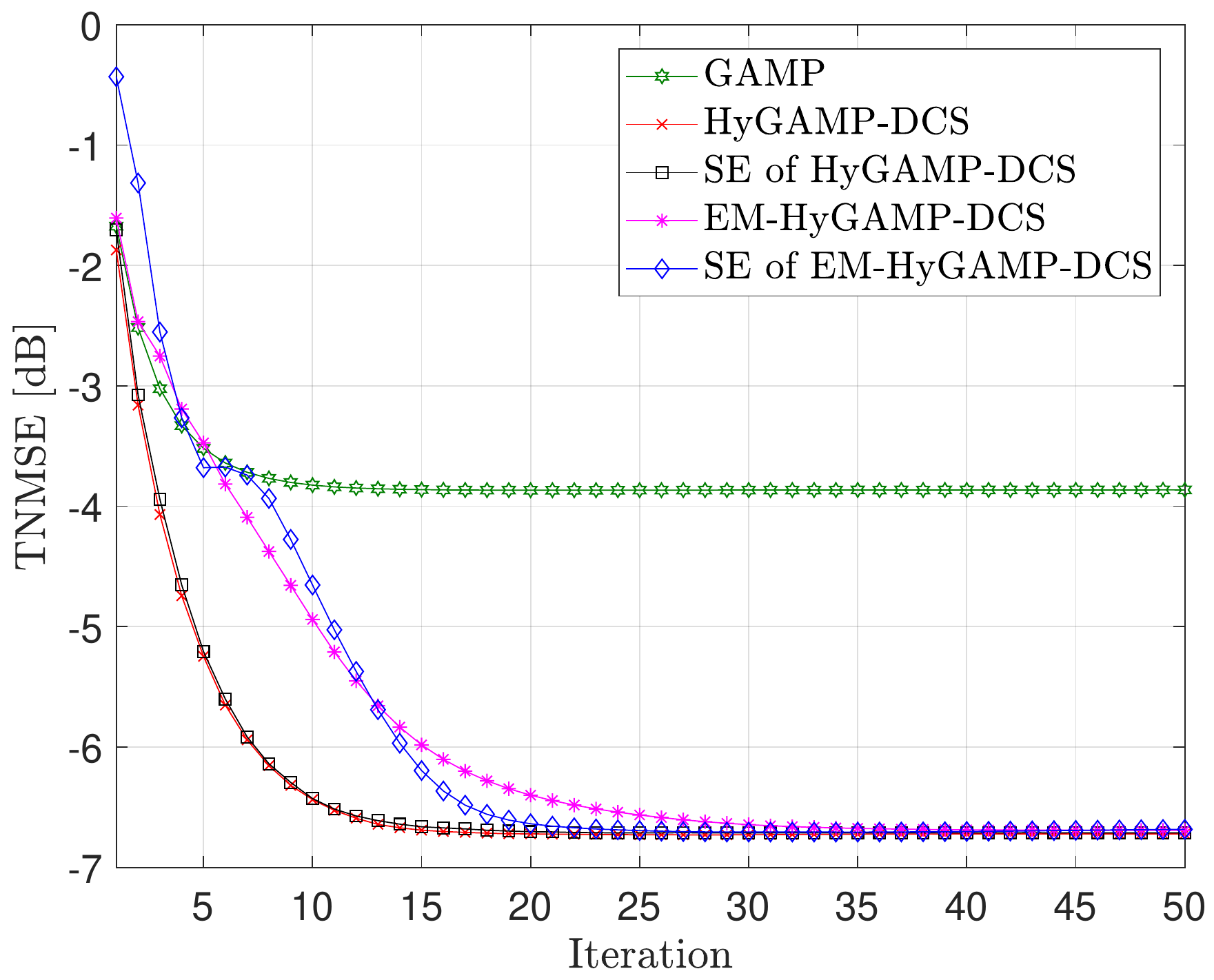}
    \vspace{-1.4cm}
    \caption{TNMSE performance with $\text{SNR} = -10 \text{dB}$.}\label{Fig:SE_-10}
  \end{minipage}
  \begin{minipage}[t]{.46\textwidth}
    \center
    \includegraphics[width=\textwidth]{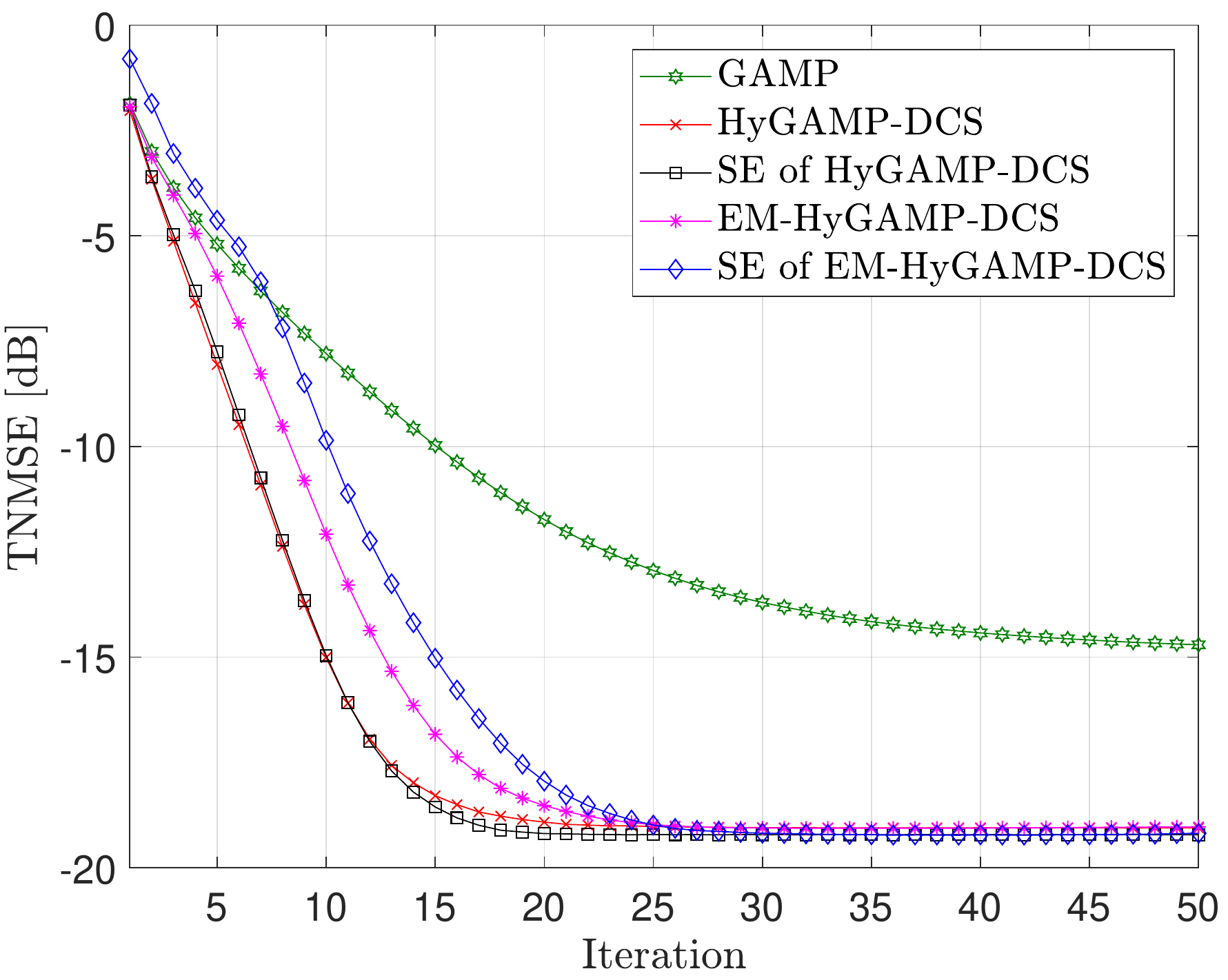}
    \vspace{-1.4cm}
    \caption{TNMSE performance with $\text{SNR} = 0 \text{dB}$.}\label{Fig:SE_0}
  \end{minipage}
  \vspace{-0.5cm}
\end{figure}

\subsection{Validation of the Theoretical Analysis}

Finally, the simulation and theoretical results under the setup of $L=200$ and $p_a=0.1$ are given in Fig. \ref{Fig:SE_-10} and Fig. \ref{Fig:SE_0}. We can find that the converged TNMSE of the proposed HyGAMP-DCS algorithm and the EM-HyGAMP-DCS algorithm can be accurately predicted by the SE. Compared with the GAMP algorithm, the proposed algorithms can have faster convergence in the high SNR scenario, which means that the HyGAMP-DCS-based algorithms can achieve performance improvement in both estimation accuracy and convergence. The theoretical results also indicate that the EM-HyGAMP-DCS algorithm can achieve almost the same performance as the HyGAMP-DCS algorithm does in the asymptotic regime even if the perfect system statistics are unknown. Therefore, in the practical system with imperfect system statistics, we can first employ the SE to find the appropriate hyperparameter initialization value and then employ the EM-HyGAMP-DCS algorithm for joint user activity detection and channel estimation.

\section{Conclusions}\label{sec:conclusion}

The grant-free temporally-correlated massive access system is studied in this work, where the active users have a large probability to transmit in multiple continuous frames. The problem of joint user activity detection and channel estimation in multiple consecutive frames is formulated as a DCS problem. Based on the probabilistic model, we propose the computationally efficient HyGAMP-DCS algorithm by exploiting both the statistics of channels and the temporally-correlated user activities from the perspective of Bayesian inference. Simulation results show that the proposed algorithm can achieve much better performance than the conventional DCS-based algorithm. To acquire the knowledge of the system statistics, the EM algorithm is proposed to be incorporated in the HyGAMP-DCS algorithm by adaptively updating the hyperparameters during the estimation procedure. In particular, we point out that the SE is the powerful tool to select the appropriate hyperparameter initialization of EM-HyGAMP-DCS. Additionally, we observe that only a moderate number of joint detection frames is enough for the proposed algorithms to nearly achieve the optimal performance.

\appendix
%\small

\subsection{{Derivation of (\ref{equ:EM_tau_w})}}
The objective function in problem (\ref{equ:M_step}) can be expressed as
\begin{align}\label{equ:ML_func}
    \mathcal{L}^i(\pmb{\vartheta}) &= \mathbb{E}_{p^i(\mathbf{X})}[\ln p(\mathbf{X},\mathbf{Y}|\pmb{\vartheta})] \notag \\
    \quad &= \mathbb{E}_{p^i(\mathbf{X})}\left[\sum_{t=1}^{T} \Big(\ln p(\mathbf{y}_t|\mathbf{x}_t) + \sum_{n=1}^{N}\ln p(x_{n,t}|\beta,\lambda_{n,t}) + \ln p(\lambda_{n,t}|\lambda_{n,t-1}) \Big)\right].
\end{align}
We can observe that all the terms except $\ln p(\mathbf{y}_t|\mathbf{x}_t)$ for all $t$ are independent to the noise variance $\sigma^2_w$, so that the partial derivative of (\ref{equ:ML_func}) with respect to $\sigma_w$ is obtained as
\begin{align}\label{equ:DML_sigma_w}
    \frac{\partial}{\partial \sigma^2_w} \mathcal{L}^i(\pmb{\vartheta}) &= \frac{\partial}{\partial \sigma^2_w} \mathbb{E}_{p^i(\mathbf{X})} \left[ \sum_{t=1}^{T} \ln p(\mathbf{y}_t|\mathbf{x}_t) \right] \notag \\
    %\quad &= \sum_{t=1}^{T} \sum_{l=1}^{L} \int p^i(z_{l,t}|\mathbf{Y}) \frac{\partial}{\partial \sigma^2_w} \ln p(y_{l,t}|z_{l,t}) dz_{l,t} \notag \\
    \quad &= \frac{1}{\sigma_w^4} \sum_{t=1}^{T}\sum_{l=1}^{L} \Big[ |y_{l,t}-\widehat{z}^0_{l,t}(i)|^2 + \tau^z_{l,t}(i) \Big] + \frac{LT}{\sigma^2_w}.
\end{align}
Finally, we can give the update rule of $\sigma^2_w$ in (\ref{equ:EM_tau_w}) by setting the equation (\ref{equ:DML_sigma_w}) equaling to zero.

\subsection{{Derivation of (\ref{equ:EM_beta})}}
Similarly, only the terms $\ln p(x_{n,t}|\beta,\lambda_{n,t})$ for all $t$ depends on the large-scale attenuation component. We can derive the partial derivative of (\ref{equ:ML_func}) with respect to $\beta$ as
\begin{align}\label{equ:DML_beta}
    \frac{\partial}{\partial \beta} \mathcal{L}^i(\pmb{\vartheta}) &=  \sum_{t=1}^{T}\sum_{l=1}^{L} \int p^i(x_{n,t}|\mathbf{Y}) \frac{\partial}{\partial \beta} \ln p(x_{l,t}|\beta,\lambda_{n,t}) dx_{n,t},
\end{align}
where
\begin{equation}\label{equ:Dpx}
    \frac{\partial}{\partial \beta} \ln p(x_{l,t}|\beta,\lambda_{n,t}) = \left\{ \begin{array}{ll}
                              0, & x_{n,t} = 0,  \\
                              \frac{|x_{n,t}|^2}{\beta^2} - \frac{1}{\beta}, & x_{n,t} \ne 0.
                            \end{array} \right.
\end{equation}

To tackle the discontinuity of $\frac{\partial p(x_{n,t}|\beta,\lambda_{n,t})}{\partial\beta}$ at the point $x_{n,t} = 0$, we adopt the strategy in \cite{Vila_2013_TSP} where the closed region $\mathcal{B} \triangleq \{x||x| \leq \epsilon, x \in \mathbb{C}\}$ and its complementary $\bar{\mathcal{B}} = \mathbb{C} \setminus \mathcal{B}$ are defined with the limit $\epsilon \to 0$. Therefore, we can obtain
\begin{align}\label{equ:Dpx_final}
    \frac{\partial}{\partial \beta} \mathcal{L}^i(\pmb{\vartheta}) &= \frac{1}{\beta^2} \sum_{t=1}^{T}\sum_{n=1}^{N} \int_{\mathcal{B}} p^i(x_{n,t}|\mathbf{Y}) |x_{n,t}^2|^2 dx_{n,t} - \frac{1}{\beta} \sum_{t=1}^{T}\sum_{n=1}^{N} \int_{\mathcal{B}} p^i(x_{n,t}|\mathbf{Y}) dx_{n,t} \notag \\
    \quad &= \frac{1}{\beta^2} \sum_{t=1}^{T}\sum_{n=1}^{N} \varpi_{n,t}(i)(|\gamma_{n,t}(i)|^2 + \tau_{n,t}^{\gamma}(i)) - \frac{1}{\beta} \sum_{t=1}^{T}\sum_{n=1}^{N} \varpi_{n,t}(i).
\end{align}
Then the update rule of $\beta$ can be derived in (\ref{equ:EM_beta}) by setting (\ref{equ:Dpx_final}) to be zero.

\subsection{{Derivation of (\ref{equ:EM_pa})}}
We can observe that only the term $\ln p(\lambda_{n,t}|\lambda_{n,t-1})$ with $t=1$ depends on $p_a$, therefore, the partial derivative of (\ref{equ:M_step}) with respect to $p_a$ is given by
\begin{align}\label{equ:DML_pa}
    \frac{\partial}{\partial p_a} \mathcal{L}^i(\pmb{\vartheta}) &= \sum_{n=1}^{N} \int p^i(\lambda_{n,1}|\mathbf{Y}) \frac{\partial}{\partial p_a} \ln p(\lambda_{n,1}|\lambda_{n,0}) d\lambda_{n,1},
\end{align}

Due to the fact that $p(\lambda_{n,1}|\lambda_{n,0})$ satisfies Bernoulli distribution, the partial derivative to $\ln p(\lambda_{n,1}|\lambda_{n,0})$ with respect to $p_a$ can be expressed as
\begin{align}\label{equ:D_p_lamb1}
    \frac{\partial}{\partial p_a} \ln p(\lambda_{n,1}|\lambda_{n,0}) &= \frac{2\lambda_{n,1}-1}{p(\lambda_{n,1}|\lambda_{n,0})} = \left\{ \begin{array}{ll}
                      -\frac{1}{1-p_a}, & \lambda_{n,1} = 0, \\
                      \frac{1}{p_a}, & \lambda_{n,1} = 1.
                    \end{array} \right.
\end{align}

From the message passing algorithm, the posterior probability of $\lambda_{n,1}$ can be obtained as
\begin{align}\label{equ:p_lamb1}
    p^i(\lambda_{n,1}|\mathbf{Y}) &= \xi^i_{(n,1) \rightarrow (n,1)}(\lambda_{n,1}) \cdot \upsilon^i_{(n,1) \leftarrow (n,1)}(\lambda_{n,1}) \cdot \zeta^i_{(n,1) \leftarrow (n,2)}(\lambda_{n,1}) \notag \\
    \quad &= (1-\kappa_{n,1}(i))(1-\lambda_{n,1}) + \kappa_{n,1}(i)\lambda_{n,1},
%    \quad &\propto (1-\overrightarrow{q}_{n,1})(1-\overleftarrow{q}_{n,1})(1-\overleftarrow{p}_{n,1})(1-\lambda_{n,1}) + \overrightarrow{q}_{n,1}\overleftarrow{q}_{n,1}\overleftarrow{p}_{n,1}\lambda_{n,1}.
\end{align}
where
\begin{align}\label{equ:kappa}
    \kappa_{n,1}(i) &= \frac{\overrightarrow{q}_{n,1}(i)\overleftarrow{q}_{n,1}(i)\overleftarrow{p}_{n,1}(i)}{(1-\overrightarrow{q}_{n,1}(i))(1-\overleftarrow{q}_{n,1}(i))(1-\overleftarrow{p}_{n,1}(i)) + \overrightarrow{q}_{n,1}(i)\overleftarrow{q}_{n,1}(i)\overleftarrow{p}_{n,1}(i)}.
\end{align}

By inserting (\ref{equ:p_lamb1}) into (\ref{equ:DML_pa}), we can express (\ref{equ:DML_pa}) as
\begin{align}\label{equ:DML_pa_final}
    \frac{\partial}{\partial p_a} \mathcal{L}^i(\pmb{\vartheta}) &= \sum_{n=1}^{N}\left[ -\frac{1-\kappa_{n,1}(i)}{1-p_a} + \frac{\kappa_{n,1}(i)}{p_a} \right].
\end{align}
Finally, we obtain (\ref{equ:EM_pa}) as the update rule for $p_a$ by setting (\ref{equ:DML_pa_final}) equal to zero.

\subsection{{Derivation of (\ref{equ:EM_p10})}}
It is found that only the terms $\ln p(\lambda_{n,t}|\lambda_{n,t-1})$s for $t \ge 2$ depend on $p_{10}$, and we have
\begin{align}\label{equ:p_trans_lamb_tt-1}
    p(\lambda_{n,t}|\lambda_{n,t-1}) =&~ (1-\lambda_{n,t-1})(1-\lambda_{n,t})p_{00} +  (1-\lambda_{n,t-1})\lambda_{n,t}p_{01} \notag \\
    \quad &+ \lambda_{n,t-1}(1-\lambda_{n,t})p_{10} + \lambda_{n,t-1}\lambda_{n,t}p_{11}.
\end{align}
Thus, the partial derivative of (\ref{equ:M_step}) with respect to $p_{10}$ is given by
\begin{align}\label{equ:DML_p10}
    \frac{\partial}{\partial p_{10}} \mathcal{L}^i(\pmb{\vartheta}) &= \sum_{t=2}^{T}\sum_{n=1}^{N} \iint p^i(\lambda_{n,t-1},\lambda_{n,t}|\mathbf{Y}) \frac{\partial}{\partial p_{10}} \ln p(\lambda_{n,t}|\lambda_{n,t-1}) d\lambda_{n,t-1}d\lambda_{n,t} \notag \\
    \quad &= \sum_{t=2}^{T}\sum_{n=1}^{N} \iint p^i(\lambda_{n,t-1},\lambda_{n,t}|\mathbf{Y}) \Bigg[ \frac{\lambda_{n,t-1}(1-\lambda_{n,t})}{p_{10}} - \frac{\lambda_{n,t-1}\lambda_{n,t}}{1-p_{10}} \Bigg] d\lambda_{n,t-1}d\lambda_{n,t} \notag \\
    \quad &=  \sum_{t=2}^{T}\sum_{n=1}^{N} \left( \frac{\mathbb{E}[\lambda_{n,t-1}]}{p_{10}} - \frac{\mathbb{E}[\lambda_{n,t-1}\lambda_{n,t}]}{p_{10}(1-p_{10})} \right).
\end{align}
Then the update rule of $p_{10}$ can be obtained as (\ref{equ:EM_p10}) by setting (\ref{equ:DML_p10}) equal to zero. To give the specific expression of the update rule for $p_{10}$, we can first obtain $\mathbb{E}[\lambda_{n,t-1}] = \kappa_{n,t-1}(i)$ in the $i$th iteration, where $\kappa_{n,t-1}(i)$ can be similarly obtain by (\ref{equ:kappa}).
%\begin{equation}\label{equ:E_lamb_t-1}
%    \mathbb{E}[\lambda_{n,t-1}] = \frac{\overrightarrow{q}_{n,t-1}(i)\overleftarrow{q}_{n,t-1}(i)\overleftarrow{p}_{n,t-1}(i)}{(1-\overrightarrow{q}_{n,t-1}(i))(1-\overleftarrow{q}_{n,t-1}(i))(1-\overleftarrow{p}_{n,t-1}(i)) + \overrightarrow{q}_{n,t-1}(i)\overleftarrow{q}_{n,t-1}(i)\overleftarrow{p}_{n,t-1}(i)}.
%\end{equation}
From the MP-based algorithm, the pairwise joint posterior probability of $ p(\lambda_{n,t-1},\lambda_{n,t}|\mathbf{Y})$ can be expressed as
\begin{align}\label{equ:p_lamb_tt-1}
    p^i(\lambda_{n,t-1},\lambda_{n,t}|\mathbf{Y}) =&~ p(\lambda_{n,t}|\lambda_{n,t-1}) \cdot \xi^i_{(n,t-1) \rightarrow (n,t)}(\lambda_{n,t-1}) \cdot \zeta^i_{(n,t-1) \leftarrow (n,t)}(\lambda_{n,t}), \notag \\
    \quad =&~ p(\lambda_{n,t}|\lambda_{n,t-1}) \cdot [(1-\phi_{n,t-1}(i))(1-\lambda_{n,t-1})+\phi_{n,t-1}(i)\lambda_{n,t-1}] \notag \\
     \quad &\cdot [(1-\varphi_{n,t}(i))(1-\lambda_{n,t})+\varphi_{n,t}(i)\lambda_{n,t}],
\end{align}
where
\begin{align}
    \phi_{n,t-1}(i) &= \frac{\overrightarrow{q}_{n,t-1}(i)\overleftarrow{p}_{n,t-1}(i)}{(1-\overrightarrow{q}_{n,t-1}(i))(1-\overleftarrow{p}_{n,t-1}(i))+\overrightarrow{q}_{n,t-1}(i)\overleftarrow{p}_{n,t-1}(i)}, \label{equ:phi_nt-1} \\
    \varphi_{n,t}(i) &= \frac{\overleftarrow{q}_{n,t}(i)\overleftarrow{p}_{n,t}(i)}{(1-\overleftarrow{q}_{n,t}(i))(1-\overleftarrow{p}_{n,t}(i))+\overleftarrow{q}_{n,t}(i)\overleftarrow{p}_{n,t}(i)}. \label{equ:varphi_nt}
\end{align}
From the joint posterior probability $p(\lambda_{n,t-1},\lambda_{n,t}|\mathbf{Y})$, we can have
\begin{align}\label{equ:E_lamb_tt-1}
    \mathbb{E}[\lambda_{n,t-1}\lambda_{n,t}] &= p_{11}\phi_{n,t-1}(i)\varphi_{n,t}(i).
\end{align}
By inserting (\ref{equ:E_lamb_tt-1}) into (\ref{equ:EM_p10}), we finally obtain the explicit expression of the update rule for $p_{10}$.
%Then the update rule of $p_{10}$ can be obtained as (\ref{equ:EM_p10}).

\bibliographystyle{IEEEtran}
\bibliography{TCMC}

\end{document}